\documentclass[journal]{IEEEtran}
\usepackage{amsfonts}

\usepackage{bm}
\usepackage{amsmath}
\usepackage[keeplastbox]{flushend}
\usepackage{amsthm}
\usepackage{amssymb}
\usepackage{makecell}
\usepackage{amssymb}
\usepackage{array}
\usepackage{cases}
\usepackage{algorithm}
\usepackage{algorithmic}
\usepackage{enumerate}
\usepackage{color,xcolor}
\usepackage{graphicx}
\usepackage{multirow,tabularx}
\usepackage{subfigure}
\usepackage{multirow}
\usepackage{hyperref}
\usepackage[compress]{cite}
\usepackage{setspace}

\newtheorem{remark}{Remark}
\newtheorem{proposition}{Proposition}
\makeatletter

\newcommand{\Rmnum}[1]{\expandafter\@slowromancap\romannumeral #1@}
\usepackage[vlined, ruled, algo2e]{algorithm2e}
\makeatother
 0

\begin{document}

\title{Asynchronous Grant-Free Random Access: Receiver Design with Partially Uni-Directional Message Passing and Interference Suppression Analysis}

\author{Zhaoji Zhang,~\IEEEmembership{Member,~IEEE,} Yuhao Chi,~\IEEEmembership{Member,~IEEE,} Qinghua Guo,~\IEEEmembership{Senior~Member,~IEEE,}\\Ying Li,~\IEEEmembership{Member,~IEEE,} Guanghui Song,~\IEEEmembership{Member,~IEEE,} and Chongwen Huang,~\IEEEmembership{Member,~IEEE}

\thanks{Zhaoji Zhang, Yuhao Chi, Ying Li, and Guanghui Song are with the School of Telecommunications Engineering, Xidian University, Xi'an 710071, China (email: zhaojizhang@xidian.edu.cn; yhchi@xidian.edu.cn; yli@mail.xidian.edu.cn; songguanghui@xidian.edu.cn).
	
Qinghua Guo is with the School of Electrical, Computer and
Telecommunications Engineering, University of Wollongong, Wollongong,
NSW 2522, Australia (e-mail: qguo@uow.edu.au)

Chongwen Huang is with the College of Information Science and Electronic Engineering, Zhejiang University, Hangzhou 310027, China, also with the International Joint Innovation Center, Zhejiang University, Haining 314400, China, and also with the Zhejiang Provincial Key Laboratory of Information Processing, Communication and Networking (IPCAN), Hangzhou 310027, China (e-mail: chongwenhuang@zju.edu.cn).
}}
\maketitle{
\begin{abstract}
Massive Machine-Type Communications (mMTC) features a massive number of low-cost user equipments (UEs) with sparse activity. Tailor-made for these features, grant-free random access (GF-RA) serves as an efficient access solution for mMTC. However, most existing GF-RA schemes rely on strict synchronization, which incurs excessive coordination burden for the low-cost UEs. In this work, we propose a receiver design for asynchronous GF-RA, and address the joint user activity detection (UAD) and channel estimation (CE) problem in the presence of asynchronization-induced inter-symbol interference. Specifically, the delay profile is exploited at the receiver to distinguish different UEs. However, a sample correlation problem in this receiver design impedes the factorization of the joint likelihood function, which complicates the UAD and CE problem. To address this correlation problem, we design a partially uni-directional (PUD) factor graph representation for the joint likelihood function. Building on this PUD factor graph, we further propose a PUD message passing based sparse Bayesian learning (SBL) algorithm for asynchronous UAD and CE (PUDMP-SBL-aUADCE). Our theoretical analysis shows that the PUDMP-SBL-aUADCE algorithm exhibits higher signal-to-interference-and-noise ratio (SINR) in the asynchronous case than in the synchronous case, i.e., the proposed receiver design can exploit asynchronization to suppress multi-user interference. In addition, considering potential timing error from the low-cost UEs, we investigate the impacts of imperfect delay profile, and reveal the advantages of adopting the SBL method in this case. Finally, extensive simulation results are provided to demonstrate the performance of the PUDMP-SBL-aUADCE algorithm.
\end{abstract}
\begin{IEEEkeywords}
Grant-free random access, asynchronization, partially uni-directional message passing, user activity detection, channel estimation
\end{IEEEkeywords}
\IEEEpeerreviewmaketitle
\section{Introduction}


\IEEEPARstart{M}assive Machine-Type Communications (mMTC) serves as a key application scenario in the fifth-generation (5G) to support diversified Internet-of-Things (IoT) applications \cite{mmtc}, such as smart cities, smart home, smart factories, etc. Unlike conventional human-oriented application scenarios, mMTC is characterized by the following distinctive features, i.e., the massiveness of potential user equipments (UEs), the sparse activity of each UE, short data packets from active UEs, as well as the need for low power consumption from the low-cost devices. Furthermore, these features will become even more prominent with the evolution of mMTC in future Beyond 5G (B5G) and 6G systems. On the other hand, in the random access channel (RACH) procedure, each active UE needs to establish a connection with the base station (BS) before data transmission. However, these above-mentioned features of mMTC will bring new challenges to the conventional RACH procedure, such as the shortage of uplink resources, random access collisions, etc. 

To address the above-mentioned challenges, the grant-free random access (GF-RA) mechanism \cite{GFRA}, also termed as grant-free non-orthogonal multiple access (GF-NOMA), has recently emerged as a promising RACH solution for mMTC. Specifically, in the GF-RA mechanism, active UEs can directly transmit their payload data, without applying for the scheduling grant to establish a connection with the BS. In this way, the GF-RA mechanism can skip the handshaking procedure of the conventional grant-based random access (GB-RA) mechanism, which reduces signaling overhead and improves the data-transmission efficiency. In addition, enabled by different NOMA techniques, the GF-RA mechanism also facilitates the sharing of access resources among all the active UEs, which improves the resource utilization efficiency and mitigates the collision problem. Since only a small amount of control signaling is exchanged between the BS and active UEs in GF-RA, several critical problems should be addressed at the BS. Specifically, the BS needs to identify the active UEs, estimate the channel state information (CSI) from active UEs, and finally detect their data. These problems are termed as user-activity detection (UAD), channel estimation (CE), and multi-user data detection (MUD), respectively. Existing solutions to these problems are briefed as follows.

In some typical GF-RA schemes \cite{YuWeiAMP,Sat,ZYYMP,DNNSBL}, each active UE transmits its unique pilot sequence along with its payload data. In this way, the UAD and CE problems are firstly addressed at the receiver, then the UAD and CE result will be subsequently used to solve the MUD problem. Furthermore, several novel schemes \cite{BianXinyu,JUICESD,VBI} have been recently proposed to jointly address the UAD, CE, and MUD problems at the receiver, so as to further improve the UAD and MUD performance. Despite that different receiver designs and detection algorithms have been proposed for GF-RA, most of existing works are built on the assumption of ideally synchronized signal reception at the BS. In other words, the data frame and data symbols from different UEs are assumed to reach the BS at exactly the same time. However, this assumption of ideal synchronization entails frequent coordination between the BS and UEs, which significantly increases the coordination burden, and contradicts the low power consumption requirement of the low-cost UEs. Confronted with this problem, the receiver design for asynchronous GF-RA has attracted increasing research interest.

\subsection{Literature Review}
The orthogonal frequency-division multiplexing (OFDM) technique can employ the cyclic prefix (CP) to handle the interference caused by different transmission delay. Therefore, different asynchronous GF-RA schemes \cite{KoreanTWC,offset,yury,TIT,TSIC} have been proposed under the OFDM-based NOMA framework. In \cite{offset,yury,TIT}, the CP is used not only to handle the multi-path delay, but also to compensate for the asynchronization between different users. However, this CP-based strategy in \cite{offset,yury,TIT} requires that the CP should be excessively long to accommodate the large delay among all the UEs. In \cite{TSIC}, the CP is only used to handle the multi-path delay, so that the required CP length can be significantly reduced. A triangular successive interference cancellation (T-SIC) algorithm was then proposed in \cite{TSIC} to tackle the interference caused by asynchronization. However, this T-SIC algorithm requires known user activity and CSI, as well as sufficiently large power difference among different users to facilitate SIC decoding. However, these assumptions may become impractical for mMTC with densely deployed and randomly activated UEs. In addition, a common problem with the OFDM-based GF-RA schemes \cite{KoreanTWC,offset,yury,TIT,TSIC} is that the OFDM technique incurs high peak-to-average power ratio (PAPR), which poses challenges for hardware implementations at the low-cost mMTC devices.

Apart from the OFDM technique, asynchronous GF-RA schemes have also been extensively studied with other NOMA-enabling techniques, such as the interleaver-division multiple access (IDMA) \cite{KoreanIDMA,LiuJingmin}, code-division multiple access (CDMA) \cite{CDMA1,CDMA2}, code-domain NOMA \cite{code1,code2,code3}, and multiple-input multiple-output (MIMO) \cite{MIMO1,MIMO2,MIMO3,MIMO4,BayesAsyn}. We further categorize these asynchronous GF-RA schemes according to different levels of asynchronization, i.e. asynchronous frame with symbol synchronization \cite{CDMA1,CDMA2,code2,code3,MIMO1,MIMO2,MIMO3,MIMO4} and asynchronization within symbol duration \cite{KoreanIDMA,LiuJingmin,code1,BayesAsyn}. The details are explained as follows. 

The GF-RA schemes in \cite{CDMA1,CDMA2,code2,code3,MIMO1,MIMO2,MIMO3,MIMO4} assume that data frames from different users are not synchronized at the receiver, yet the sampling at the BS is \emph{ideally} aligned with the symbol interval, so that symbol-level synchronization can be achieved. Under this assumption, compressed-sensing based algorithms have been proposed for UAD and CE \cite{code3} or joint delay learning (DL), UAD and MUD \cite{CDMA1,CDMA2}, while a learned approximate message passing (LAMP) algorithm was proposed for joint DL, UAD and CE \cite{code2}. In the MIMO system, a generalized AMP (GAMP) algorithm was proposed for joint UAD and CE \cite{MIMO4}, a block coordinate descent algorithm \cite{MIMO2} and a LAMP algorithm \cite{MIMO3} were proposed for the joint DL, UAD and CE problem, while a Bilinear GAMP (BiGAMP) algorithm \cite{MIMO4} was further proposed for joint DL, UAD, CE, and MUD. However, the assumption of symbol synchronization in \cite{CDMA1,CDMA2,code2,code3,MIMO1,MIMO2,MIMO3,MIMO4}, i.e. sampling perfectly aligned with the symbol interval, can be hard to realize due to the lack of coordination among the massive UEs.

In contrast to above-mentioned schemes, the asynchronous GF-RA schemes in \cite{KoreanIDMA,LiuJingmin,code1,BayesAsyn} dropped the infeasible assumption of symbol-level synchronization, and consider the asynchronization within each symbol duration. A transceiver structure was proposed in \cite{KoreanIDMA} with auxiliary preamble designed for UAD, and a cross-symbol message passing MUD algorithm was proposed in \cite{LiuJingmin}, while a block AMP algorithm was proposed in \cite{code1} for the CE and MUD problem. However, \emph{ideal knowledge} on user activity \cite{LiuJingmin,code1} or CSI \cite{LiuJingmin,KoreanIDMA} is still required in these schemes. Most recently, a turbo AMP (TAMP) algorithm has been designed in \cite{BayesAsyn} for the joint DL, CE, and MUD problem. However, this algorithm is designed for the novel \emph{unsourced} massive access (UMA) scenario, where the BS is concerned with specific payload data instead of the identity of each active UE. In other words, the UAD problem is \emph{bypassed}, but the \emph{exact number of active UEs} is still required in \cite{BayesAsyn}, while this assumption may become infeasible for the sourced random access scenario considered in  \cite{KoreanTWC,offset,yury,TIT,TSIC,KoreanIDMA,LiuJingmin,CDMA1,CDMA2,code1,code2,code3,MIMO1,MIMO2,MIMO3,MIMO4} and this paper.

\subsection{Motivations and Contributions}
According to the literature review, it is concluded that existing asynchronous GF-RA schemes are still subject to several limitations, such as the infeasible assumption of symbol-level synchronization or ideal knowledge of the user activity or CSI, etc. On the other hand, mMTC devices are commonly deployed at fixed locations with no or low mobility \cite{fix1,fix2,fix3}, so that the BS can potentially obtain the delay profile of the UEs\footnote{The assumption of available delay profile is commonly adopted in \cite{TSIC,LiuJingmin,code1}. UEs can transmit preamble sequences to facilitate estimation of the delay profile at the BS, which is a common practice in LTE-based RACH design \cite{TA}. Since mMTC devices are commonly deployed with no or low mobility \cite{fix1,fix2,fix3}, the transmission delay of each UE remains constant in a long time. Therefore, acquiring delay profile at the BS will not incur excessive signaling overhead or frequent coordination for the low-cost UEs.}. Therefore, we are motivated to exploit this delay profile to distinguish different UEs at the receiver. Furthermore, confronted with the massiveness of UE and the complicated inter-symbol interference caused by symbol asynchronization, it is desirable to develop a low-complexity detection algorithm for the joint UAD and CE problem in asynchronous GF-RA. With the UAD and CE result, the subsequent MUD problem can be readily addressed with existing solutions \cite{LiuJingmin}.    

In this paper, we propose a receiver design for asynchronous GF-RA, where the delay profile (also called as delay information) is exploited at the BS to distinguish different UEs. In the presence of asynchronization-induced inter-symbol interference, we propose a partially uni-directional (PUD) message passing based sparse Bayesian learning (SBL) algorithm for asynchronous UAD and CE (PUDMP-SBL-aUADCE). Our main contributions are summarized as follows.
\begin{figure*}
	\centering
	\includegraphics[width=2\columnwidth]{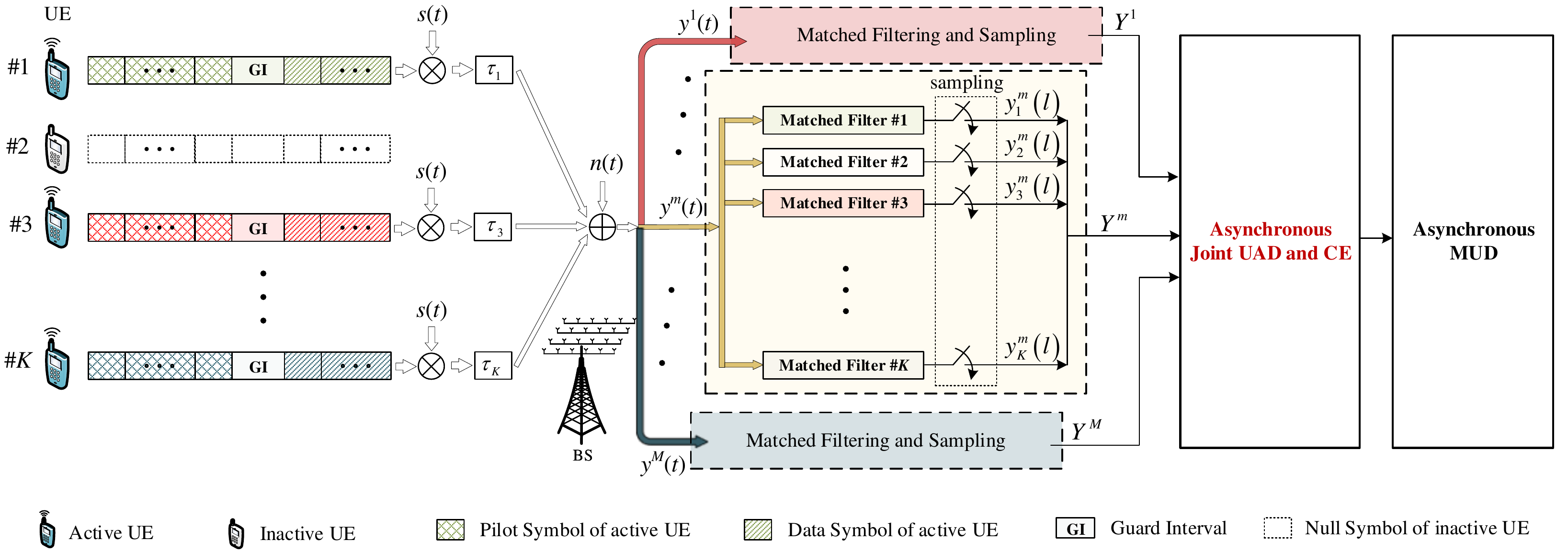}
	\caption{Asynchronous grant-free random access system model.}
	\label{sys_mod}
\end{figure*}

(i) A receiver design is proposed where the delay profile is exploited to distinguish different UEs. In order to address the sample correlation problem in this receiver design, a PUD factor graph is established. Based on this PUD factor graph, a low-complexity message passing based solution, i.e. the PUDMP-SBL-aUADCE algorithm is proposed to address the joint UAD and CE problem for asynchronous GF-RA.

(ii) The message passing procedure is analyzed, which proves that the PUDMP-SBL-aUADCE algorithm exhibits higher signal-to-interference-and-noise ratio (SINR) in the asynchronous case than in the synchronous case. In other words, the proposed receiver can effectively exploit asynchronization to suppress multi-user interference (MUI), which improves the UAD and CE accuracy. 

(iii) Considering the potential timing error resulting from the low-cost oscillators or the mobility of the devices, we investigate the impacts of imperfect delay profile at the BS, and further explain the advantages of adopting the SBL method in the proposed PUDMP-SBL-aUADCE algorithm.

The rest of this paper is organized as follows. The system model and receiver design are introduced in Section  \ref{System_section}. The proposed PUDMP-SBL-aUADCE algorithm is explained in Section \ref{AUADCE}, together with the complexity analysis, SINR analysis, as well as the discussion on the impacts of imperfect delay information. Simulation results are provided in Section \ref{simu_section}, and the paper is finally concluded in Section \ref{conclusion}.
 
\emph{Notations:} Scalars and vectors are written in italic letters and boldface lower-case letters, respectively, while matrices are in boldface upper-case letters. All the vectors are column vectors, and $(\cdot)^T$ is the transpose operation. $\mathbb{E}[\cdot]$ and $\mathbb{V}[\cdot]$ take the expectation and variance of a random variable, respectively. $X\sim\mathcal{CN}(\mu,v)$ and $X\sim\mathcal{N}(\mu,v)$ represent that a random variable $X$ follows a complex Gaussian distribution and a real Gaussian distribution with mean $\mu$ and variance $v$, respectively. $Gam(\gamma;\epsilon,\eta)$ and $\mathcal{CN}(x|\mu,v)$ denote the probability density function (pdf) of a Gamma distribution and a complex Gaussian distribution, respectively. The symbol $\propto$ means ``proportional to''. $\delta(x)$ is the Dirac function with $\delta(x)=0$ for $\forall x\neq0$, and $\int_{-\infty}^{\infty}\delta(x)dx=1$. In this paper, the terms UE, device, and user are used interchangeably.

\section{System Model}\label{System_section}
\begin{figure}
	\centering
	\includegraphics[width=1\columnwidth]{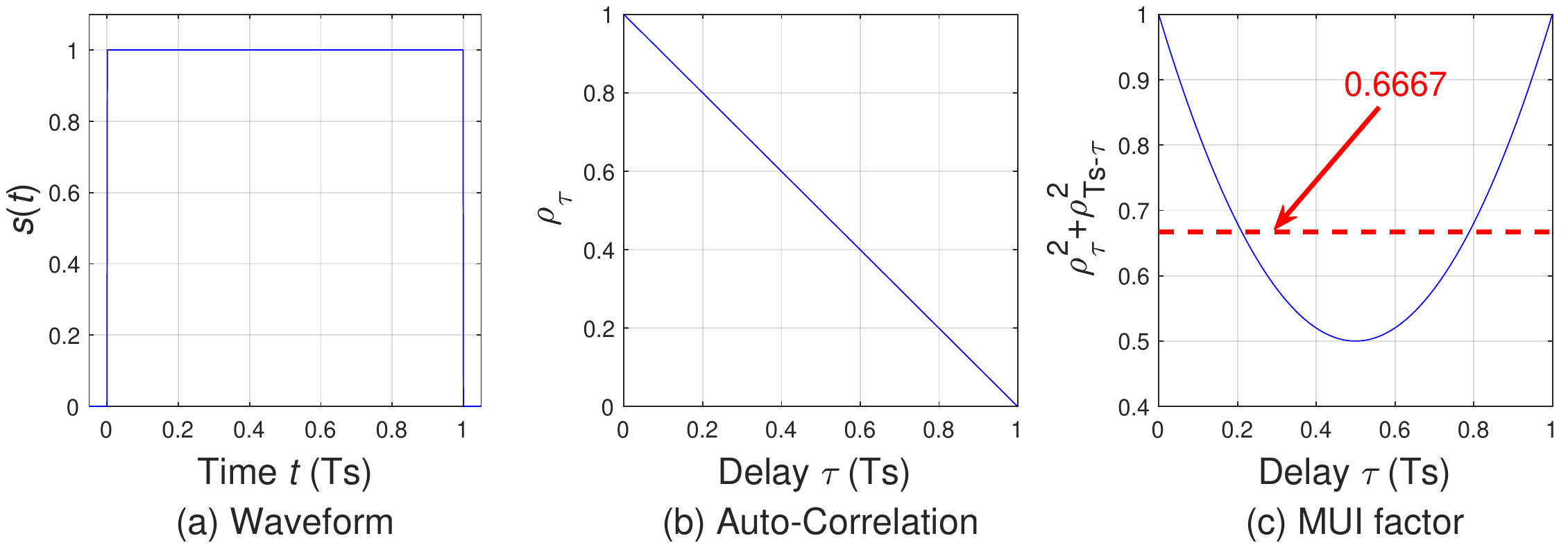}
	\caption{(a) Signal waveform $s(t)$ with (b) its auto-correlation $\rho_\tau$, and (c) the MUI factor $\rho^2_\tau+\rho^2_{T_s-\tau}$ of a rectangular waveform.}
	\label{waveform}
\end{figure}
\subsection{Transmitter Side}
As shown in Fig. \ref{sys_mod}, we consider an asynchronous cellular system, where one BS serves $K$ potential UEs. The BS is equipped with $M$ antennas, while each UE is equipped with one antenna. We further denote $\mathcal{K}=\{1,2,\ldots,K\}$ and $\mathcal{M}=\{1,2,\ldots,M\}$ as the UE-index set and antenna-index set, respectively. In each round of random access, we assume that each UE is independently activated with a small probability $p_a$, and the active UEs transmit their payload data in a grant-free manner. Specifically, if the $k$-th UE is activated, it will transmit its user-specific pilot sequence $\bm{p}_k=[p^1_k,\ldots,p_k^{L_p}]^T$, followed by a guard interval (GI) and its payload data sequence $\bm{d}_k=[d^1_k,\ldots,d_k^{L_d}]^T$, where $L_p$ and $L_d$ denote the pilot length and the data length, respectively. The set of pilot-symbol indexes is denoted as $\mathcal{L}_p=\{1,2,\ldots,L_p\}$. 


Before carrier modulation, each pilot symbol and data symbol is passed through a shaping filter \cite{LiuJingmin,BayesAsyn}, which is characterized by a pulse-shaping signal waveform $s(t)$. We note that the proposed method in this paper does not rely on any particular pulse shape, but for simplicity, we assume a rectangular waveform $s(t)$ in Fig. \ref{waveform}(a), which has unit energy during the symbol period $T_s$, i.e. $\int_{0}^{T_s}s^2(t)dt=1$, and becomes zero outside the interval $[0,T_s]$. The auto-correlation $\rho_{\tau}$ and the MUI factor $\rho^2_{\tau}+\rho^2_{T_s-\tau}$ of this rectangular waveform $s(t)$ are plotted in Fig. \ref{waveform}(b) and Fig. \ref{waveform}(c) respectively, and they will be later explained in the following sections. The waveform for other commonly-used shaping filters can be found in \cite{LiuJingmin}, and therefore not presented here for brevity. In this way, the baseband waveform signal for the $l$-th pilot symbol $p_k^l$ of the $k$-th UE is obtained as $p_k^ls\big(t-(l-1)T_s\big)$. Then, the baseband waveform signal is modulated, and transmitted to the BS.

Denote $\tau_k$ as the transmission delay of the $k$-th UE, which is mainly determined by the distance from the $k$-th UE to the BS. We do not perform synchronization to compensate for different transmission delay, which relieves the coordination burden for the low-cost UEs. Furthermore, we will later show in Section \ref{SINRana} that the proposed receiver design can effectively exploit the asynchronization to suppress MUI. It is assumed that the delay information $\{\tau_k, k\in\mathcal{K}\}$ can be obtained by the BS, since mMTC devices are mostly deployed at fixed locations with no or low mobility \cite{fix1,fix2,fix3}. Still, we will also investigate the impacts of imperfect delay information in Section \ref{nonidealTA}, which may be potentially caused by device mobility or the timing error of the low-cost local oscillators. In addition, the maximum relative delay between any two UEs is defined as $\Delta^{\max}_{\tau}=\max\limits_{\forall k,k^\prime\in\mathcal{K}}|\tau_k-\tau_{k^\prime}|$. Then, the length of the guard interval (GI) should be configured to be larger than $\Delta^{\max}_{\tau}$, so that the received pilot signal will not overlap with the received data signal.


\subsection{Receiver Side}\label{rcvsd}
The delay profile $\{\tau_k, k\in\mathcal{K}\}$ is exploited at the receiver to distinguish different UEs. Specifically, the matched filtering operation and sampling operation are performed at the BS on each antenna, and each matched filter is aligned with the delay $\tau_k$ of one UE $k\in\mathcal{K}$.  

Take the $m$-th antenna as an example, and denote $y^m(t)$ as the received baseband waveform signal, which is written as \cite{LiuJingmin,BayesAsyn}
\begin{equation}\label{time_rcv}
y^m(t)=\sum\limits^{K}_{k=1}\sum\limits^{L_p}_{l=1}\alpha_kg^m_kp^l_ks\Big(t-(l-1)T_s-\tau_k\Big) + n^m(t),
\end{equation}
where $g^m_k=\bar{g}^m_k\exp^{-j2\pi f_c \tau_k}$ is the baseband equivalent channel gain \cite{BayesAsyn}, $\bar{g}^m_k\sim\mathcal{CN}(0,1)$ is the channel coefficient from the $k$-th UE to the $m$-th antenna, $f_c$ represents the carrier modulation frequency, $n^m(t)$ denotes the additive white Gaussian noise (AWGN), $\alpha_k$ is the activity indicator, i.e. $\alpha_k=1$ if the $k$-th UE is activated, and $\alpha_k=0$ otherwise. In addition, we assume a block fading channel, i.e., the channel gains $\{g^m_k, m\in\mathcal{M}, k\in\mathcal{K}\}$ remain constant during each round of random access.

For notation simplicity, the UEs are indexed by the value of the transmission delay, i.e. $\tau_{k^\prime}\leq\tau_k$ for $\forall k^\prime<k$, and the maximum relative delay is $\Delta^{\max}_{\tau}=\tau_K-\tau_1$. Since we mainly focus on the asynchronization within each symbol duration, it is assumed that the maximum relative delay is shorter than one symbol period, i.e. $\Delta^{\max}_{\tau}<T_s$. This assumption does not sacrifice any generality, and we take a two-UE case with $\tau_1=0.2T_s$ and $\tau_2=1.5T_s$ as an example for explanation. In this case, according to (\ref{time_rcv}), the $l$-th pilot symbol of UE 1 is overlapped with the $(l-2)$-th and $(l-1)$-th pilot symbols of UE 2 at the receiver side. Simply increasing the pilot index of UE 2 by one, we have the effective delay $\tau^\text{eff}_2=\tau_2-T_s=0.5T_s$ for UE 2. In this way, the assumption that $\Delta^{\max}_{\tau}<T_s$ can always hold for any UE number $K$. 

The received pilot waveform signal $y^m(t)$ is passed through $K$ user-oriented matched filters in parallel, where the $k$-th matched filter is aligned with the delay $\tau_k$ of the $k$-th UE. Then, denote $y^m_k(l)$ as the $l$-th sampled output of the $k$-th matched filter on the $m$-th antenna, and we have
\begin{equation}\label{digital_rcv}
\begin{split}
&y^m_k(l)\overset{(a)}{=}\int^{lT_s+\tau_k}_{(l-1)T_s+\tau_k}y^m(t)s\Big(t-(l-1)T_s-\tau_k\Big)dt,\\
&\overset{(b)}{=}\alpha_kg^m_kp^l_k+\sum\limits_{k^\prime<k}\alpha_{k^\prime}{g}^m_{k^\prime}\Big(p^l_{k^\prime}\rho_{\tau_k-\tau_{k^\prime}}+p^{l+1}_{k^\prime}\rho_{T_s-(\tau_k-\tau_{k^\prime})}\Big)\\
&\ \ \ +\sum\limits_{k^\prime>k}\alpha_{k^\prime}{g}^m_{k^\prime}\Big(p^{l-1}_{k^\prime}\rho_{\tau_{k^\prime}-\tau_k}+p^{l}_{k^\prime}\rho_{T_s-(\tau_{k^\prime}-\tau_k)}\Big)+n^m_k(l),\\
&\overset{(c)}{=}\sum\limits^K_{k^\prime=1}\alpha_{k^\prime}g^m_{k^\prime}\bar{p}^l_{k,k^\prime}+n^m_k(l),\\
\end{split}
\end{equation}  
where equation ($b$) of (\ref{digital_rcv}) is obtained by substituting (\ref{time_rcv}) into equation ($a$), and $n^m_k(l)$ is the noise sample with distribution $n^m_k(l)\sim\mathcal{CN}(0,\sigma^2_n)$. Define the \emph{auto-correlation} $\rho_\tau$ of $s(t)$ as $\rho_\tau\overset{\Delta}{=}\int_{0}^{T_s}s(t)s(t-\tau)dt$, which has the property $\rho_{-\tau}=\rho_\tau$ by definition. It is shown that $\rho_\tau$ is a function of the delay $\tau$, and $\rho_\tau$ for the rectangular waveform is plotted in Fig. \ref{waveform}(b). Furthermore, we denote $\bar{p}^l_{k,k^\prime}$ as the $l$-th \emph{effective} pilot symbol of the $k^\prime$-th UE at the $k$-th matched filter, so that the notations can be unified as in equation ($c$) of (\ref{digital_rcv}), and $\bar{p}^l_{k,k^\prime}$ is written as
\begin{equation}\label{eff_pilot}
\bar{p}^l_{k,k^\prime}{=}\left\{ \begin{array}{l}
p^l_k,\ \ \ \ \ \ \ \ \ \ \ \ \ \ \ \ \ \ \ \ \ \ \ \ \ \ \ \ \ \ \ \ \ \ \ \text{if}\ k^\prime=k, \\
p^l_{k^\prime}\rho_{\tau_k-\tau_{k^\prime}}+p^{l+1}_{k^\prime}\rho_{T_s-(\tau_k-\tau_{k^\prime})},\ \ \text{if}\ k^\prime<k,\\
p^{l-1}_{k^\prime}\rho_{\tau_{k^\prime}-\tau_k}+p^{l}_{k^\prime}\rho_{T_s-(\tau_{k^\prime}-\tau_k)},\,\,  \ \text{if}\ k^\prime>k.
\end{array} \right.
\end{equation}
\begin{figure}
	\centering
	\includegraphics[width=1\columnwidth]{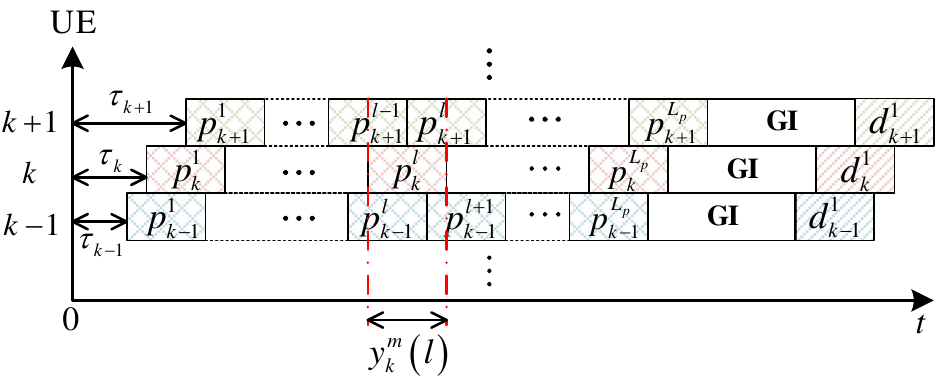}
	\caption{Multi-user inter-symbol interference caused by asynchronization.}
	\label{MUI}
\end{figure}

As shown in equation ($b$) of (\ref{digital_rcv}) and Fig. \ref{MUI}, the $l$-th pilot symbol $p^l_k$ of UE $k$ is interfered by two adjacent pilot symbols from another UE $k^\prime$ due to the symbol asynchronization. If all the UEs share the same delay, $y^m_k(l)$ in (\ref{digital_rcv}) will reduce to the symbol-synchronized case, i.e. $y^m_k(l)=\sum^K_{k=1}\alpha_{k}g^m_{k}p^l_k+n^m_k(l)$, since we have $\rho_0=1$ and $\rho_{T_s}=0$.

After the matched filtering and sampling operations on each antenna, we can get the expression of the discrete received pilot signal in (\ref{digital_rcv}), which facilitates subsequent UAD and CE. 
\section{Partially Uni-Directional Message Passing Based Sparse Bayesian Learning for Asynchronous UAD and CE}\label{AUADCE}
\subsection{Problem Formulation}\label{probform}
To facilitate joint UAD and CE at the BS, we first define the \emph{effective channel gain} $h_k^m$ as $h_k^m\overset{\Delta}{=}\alpha_kg_k^m$ from the $k$-th UE to the $m$-th antenna. Furthermore, we organize all the effective channel gains in a matrix $\bm{H}=[\bm{h}_1,\ldots,\bm{h}_K]^T$ with size $K\times M$, where $\bm{h}_k=[h^1_k,\ldots,h_k^m,\ldots,h^M_k]^T$ is the effective channel-gain vector of UE $k$. We can also write $\bm{H}=[\bm{h}^1,\ldots,\bm{h}^m,\ldots,\bm{h}^M]$ with $\bm{h}^m=[h^m_1,\ldots,h^m_K]^T$. It is noted that $\alpha_k=0$ for inactive UEs, and the matrix $\bm{H}$ therefore exhibits the \emph{row-wise} sparsity. If the $k$-th UE is inactive, we have $\bm{h}^T_k=\bm{0}_{1\times M}$ for the $k$-th row of $\bm{H}$. If the $k$-th UE is active, we have $\bm{h}^T_k\sim\mathcal{CN}(\bm{0}_{1\times M},\bm{I}_M)$, where $\bm{I}_M$ is an identity matrix with size $M\times M$, indicating that each element of $\bm{h}_k$ is independently and identically distributed (\emph{i.i.d}). According to the row-wise sparsity of $\bm{H}$, we follow the sparse Bayesian learning (SBL) method, and adopt a hierarchical prior probability for $\bm{H}$, i.e.
\begin{equation}\label{threelayer}
p(\bm{H})=\int_{\bm{\gamma}}p(\bm{H}|\bm{\gamma})p(\bm{\gamma}) d\bm{\gamma},
\end{equation}    
where $p(\bm{H}|\bm{\gamma})$ is the conditional prior pdf of $\bm{H}$, $p(\bm{\gamma})$ is the prior pdf of  $\bm{\gamma}=[\gamma_1,\ldots,\gamma_k,\ldots,\gamma_K]^T$, and $\gamma_k$ is the activity-related hyper-parameter for the $k$-th UE.

We further organize all the received pilot signals on the $m$-th antenna in a matrix $\bm{Y}^m=[\bm{y}^m_1,\ldots,\bm{y}^m_k,\ldots,\bm{y}^m_K]$ with size $L_p\times K$, where $\bm{y}^m_k=[{y}^m_k(1),\ldots,{y}^m_k(l),\ldots,{y}^m_k(L_p)]^T$ is the sampled output from the $k$-th matched filter on the $m$-th antenna, and $y^m_k(l)$ is given in (\ref{digital_rcv}). Then, the collection of $\bm{Y}^m$ on all the antennas is denoted as $\{\bm{Y}^m,m\in\mathcal{M}\}$. Furthermore, we denote $\sigma^2_n$ as the variance of the noise sample $n^m_k(l)$ in (\ref{digital_rcv}), and $\lambda=1/\sigma^2_n$ as the noise precision. Both $\sigma^2_n$ and $\lambda$ are assumed unknown to the BS, while we assume that the prior pdf $p(\lambda)$ follows $p(\lambda)\propto1/\lambda$. In this way, the joint a posteriori pdf of $\bm{H}$, $\bm{\gamma}$, and $\lambda$ is written as
\begin{equation}\label{APP}
\begin{split}
&p\Big(\bm{H},\bm{\gamma},\lambda|\{\bm{Y}^m, m\in\mathcal{M}\}\Big)\\
&\propto p\Big(\{\bm{Y}^m,m\in\mathcal{M}\}|\bm{H},\lambda\Big)p(\bm{H}|\bm{\gamma})p(\bm{\gamma})p(\lambda),\\
&\overset{(d)}{=}p(\lambda)\prod_{m=1}^{M}p\Big(\bm{Y}^m |\bm{h}^m,\lambda\Big)\prod_{k=1}^{K}\prod_{m=1}^{M}p(h^m_k|\gamma_k)\prod_{k=1}^{K}p(\gamma_k),
\end{split}
\end{equation}
where equation ($d$) of (\ref{APP}) is obtained by the fact that the effective channel gain $\bm{h}_k$ is independent among different UE, and that the received pilot signal $\bm{Y}^m$ on different antennas are also mutually independent. It is assumed that $p(h^m_k|\gamma_k)=\mathcal{CN}(h^m_k;0,\gamma^{-1}_k)$, so that we have $\gamma_k=1$ if the $k$-th UE is active, and $\gamma_k=+\infty$ otherwise. Therefore, $\gamma_k$ is dubbed the activity-related hyper-parameter. Following the SBL method \cite{ZYYMP}, we assume $p(\gamma_k)=Gam(\gamma_k;\epsilon,\eta)$, so that the UAD problem can be solved by estimating $\gamma_k,k\in\mathcal{K}$.

The joint UAD and CE problem is equivalent to finding $\bm{H}$, i.e. detecting the nonzero channel gain vector $\bm{h}_k$ and further estimating its value. However, finding the maximum a posteriori solution of $\bm{H}$ that maximizes $p\big(\bm{H}|\{\bm{Y}^m, m\in\mathcal{M}\}\big)=\int_{\bm{\gamma},\lambda}p\big(\bm{H},\bm{\gamma},\lambda|\{\bm{Y}^m, m\in\mathcal{M}\}\big)d\bm{\gamma} d\lambda$ incurs prohibitively high computational complexity. As an alternative, we establish a factor graph representation for the factorization in (\ref{APP}), and further propose a low-complexity message passing based solution. However, we observe a \emph{sample correlation} problem in the receiver design, which impedes the factorization of the joint likelihood function $p\big(\bm{Y}^m |\bm{h}^m,\lambda\big)$ in (\ref{APP}), and therefore complicates the UAD and CE problem. To address this problem, a novel PUD factor graph is established, based on which the PUDMP-SBL-aUADCE algorithm is derived. The details are explained as follows.
\subsection{Sample Correlation Problem and PUD Factor Graph}\label{samcorpro}
We consider the term $p\big(\bm{Y}^m |\bm{h}^m,\lambda\big)$ in (\ref{APP}), which is the joint likelihood function of all the sampled outputs $\{\bm{y}^m_k, k\in\mathcal{K}\}$ from $K$ different matched filters on the $m$-th antenna. According to (\ref{digital_rcv}),  the $k$-th column $\bm{y}^m_k$ of $\bm{Y}^m$ is written as
\begin{equation}\label{rcv_vec}
\bm{y}^m_k=\bar{\bm{P}}_k\bm{h}^m+\bm{n}^m_k,
\end{equation}
where the element $\bar{p}^l_{k,k^\prime}$ on the $l$-th row and $k^\prime$-th column of $\bar{\bm{P}}_k$ is given in (\ref{eff_pilot}). It is noted that the noise samples in $\bm{n}^m_k$, i.e. $\{{n}^m_k(l), l\in\mathcal{L}_p\}$ are \emph{mutually independent}, since they are sampled from non-overlapping time intervals. Hence, we have 
\begin{equation}
p(\bm{y}^m_k |\bm{h}^m,\lambda)=\prod_{l=1}^{L_p}\mathcal{CN}\Big({y}^m_k(l);(\bar{\bm{p}}^l_k)^T\bm{h}^m,\lambda^{-1}\Big),
\end{equation}
where $(\bar{\bm{p}}^l_k)^T$ is the $l$-th row of $\bar{\bm{P}}_k$, ${y}^m_k(l)$ is the $l$-th element of $\bm{y}^m_k$. However, we cannot further factorize the joint likelihood function $p\Big(\bm{Y}^m |\bm{h}^m,\lambda\Big)$ as $\prod_{k=1}^{K}p(\bm{y}^m_k |\bm{h}^m,\lambda)$ because the sampled output from different matched filters, i.e. different columns of $\bm{Y}^m$ are correlated. Specifically, consider the example in Fig. \ref{MUI}, where the $l$-th sampled output of the $k$-th matched filter $y^m_k(l)$ is given in (\ref{digital_rcv}). The noise sample $n^m_k(l)$ in $y^m_k(l)$ is expressed as
\begin{equation}\label{noise_sample}
n^m_k(l)=\int^{lT_s+\tau_k}_{(l-1)T_s+\tau_k}n^m(t)s\Big(t-(l-1)T_s-\tau_k\Big),
\end{equation}
and we can obtain a similar expression for the noise sample $n^m_{k-1}(l)$ in $y^m_{k-1}(l)$ from the ($k-1$)-th matched filter. Since the Gaussian noise $n(t)$ is white and circularly symmetric, the correlation coefficient $\bar{\rho}[n^m_{k}(l),n^m_{k-1}(l)]$ between $n^m_{k}(l)$ and $n^m_{k-1}(l)$, and the correlation coefficient $\bar{\rho}[n^m_{k}(l),n^m_{k-1}(l+1)]$ between $n^m_{k}(l)$ and  $n^m_{k-1}(l+1)$ are derived as
\begin{equation}\label{cor_noise_sample}
\begin{split}
&\bar{\rho}[n^m_{k}(l),n^m_{k-1}(l)]\\
&=\frac{\mathbb{E}\Big[\big(n^m_k(l)\!-\!\mathbb{E}[n^m_k(l)]\big)\big(n^{*m}_{k\!-\!1}(l)\!-\!\mathbb{E}[n^{*m}_{k\!-\!1}(l)]\big)\Big]}{\sqrt{\mathbb{V}\big(n^m_k(l)\big)\mathbb{V}\big(n^m_{k\!-\!1}(l)\big)}},\\
&=\frac{\mathbb{E}[n^m_k(l)n^{*m}_{k-1}(l)]}{\sigma_n^2}=\rho_{\tau_k-\tau_{k-1}},\\
&\bar{\rho}[n^m_{k}(l),n^m_{k-1}(l+1)]=\rho_{T_s-(\tau_k-\tau_{k-1})},
\end{split}
\end{equation}
where $n^{*m}_{k-1}(l)$ is the conjugate of $n^{m}_{k-1}(l)$. We can easily generalize (\ref{cor_noise_sample}) as $\bar{\rho}[n^m_{k}(l),n^m_{k^\prime}(l)]=\rho_{\tau_k-\tau_{k^\prime}}$ and $\bar{\rho}[n^m_{k}(l),n^m_{k^\prime}(l+1)]=\rho_{T_s-(\tau_k-\tau_{k^\prime})}$ for $\forall k^\prime<k$. It is shown in (\ref{cor_noise_sample}) that the noise samples from different matched filters are correlated. This correlation problem impedes further factorization of the joint likelihood function, i.e. $p\big(\bm{Y}^m |\bm{h}^m,\lambda\big)\neq\prod_{k=1}^{K}p(\bm{y}^m_k |\bm{h}^m,\lambda)$, which leads to a challenge in designing low-complexity message passing algorithms. 

In order to facilitate message passing based solution of the joint UAD and CE problem, we need to establish a factor graph representation for the factorization in (\ref{APP}). Messages passed on the factor graph are assumed to be independent. However, the sampled output from different matched filters are actually correlated as explained above, and this correlation problem will drastically undermine the convergence of the message passing algorithm. To address this correlation problem, we design a partially uni-directional (PUD) factor graph representation for the joint likelihood function $p\big(\bm{Y}^m |\bm{h}^m,\lambda\big)$ as in Fig. \ref{PUD}.
\begin{figure}
	\centering
	\includegraphics[width=1\columnwidth]{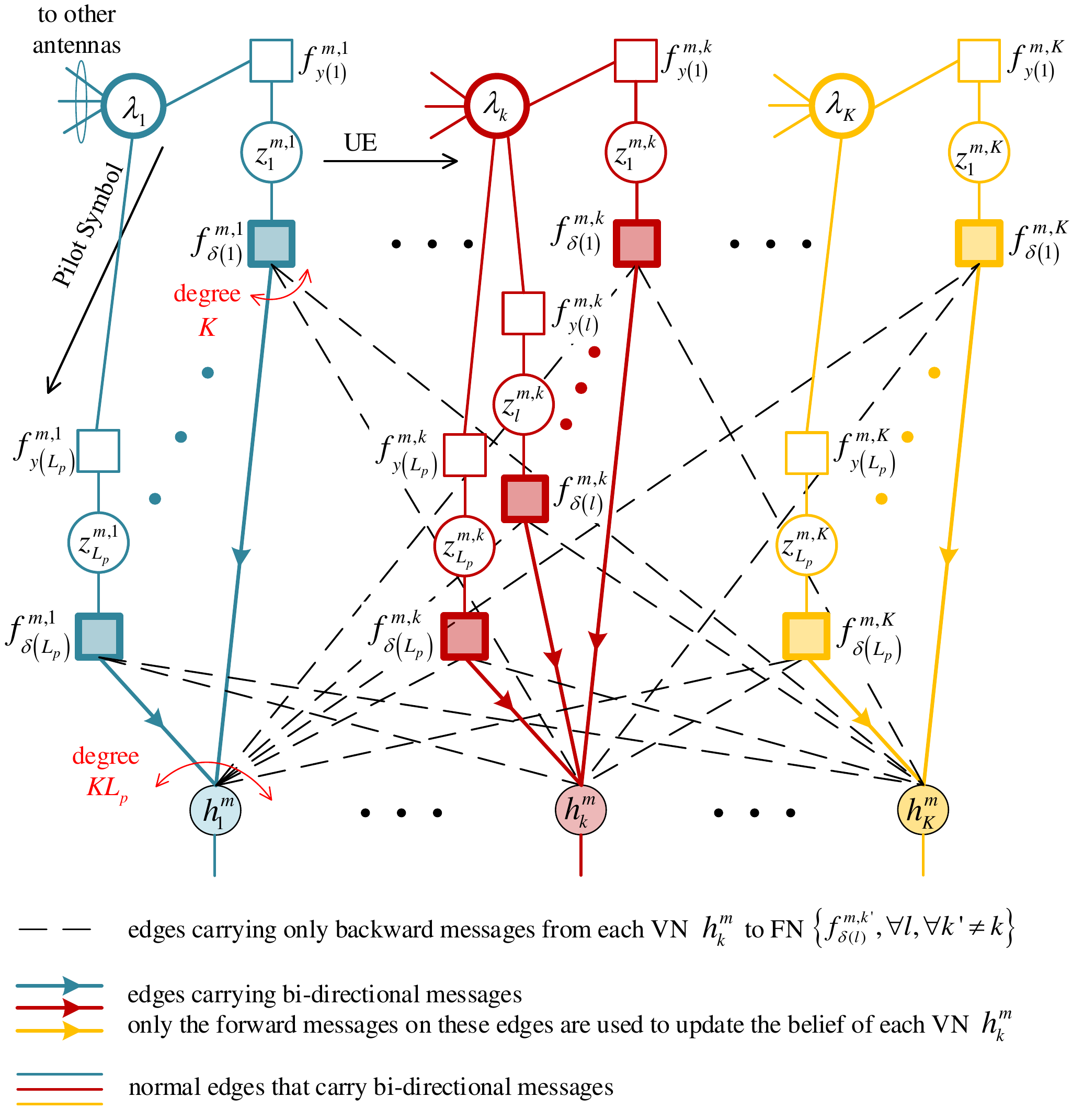}
	\caption{Partially uni-directional factor graph for the $m$-th antenna.}
	\label{PUD}
\end{figure}

The PUD factor graph in Fig. \ref{PUD} is established for the channel gain vector $\bm{h}^m=[h^m_1,\ldots,h^m_k,\ldots,h^m_K]^T$ and received signal $\bm{Y}^m$ on the $m$-th antenna. Specifically, the variable node (VN) $h^m_k$ represents the channel gain from the $k$-th UE to the $m$-th antenna. The VN $z^{m,k}_l$ is defined as the noiseless received signal  $z^{m,k}_l\overset{\Delta}{=}(\bar{\bm{p}}^l_k)^T\bm{h}^m$. Since the noise samples are mutually independent within the same matched filter but correlated between different matched filters, we set $K$ noise precision VNs $\{\lambda_k,k\in \mathcal{K}\}$, one for each matched filter. The function node (FN) $f^{m,k}_{\delta(l)}$ represents the equality constraint between two types of VNs, i.e. $f^{m,k}_{\delta(l)}(z^{m,k}_l,\bm{h}^m)=\delta\big(z^{m,k}_l-(\bar{\bm{p}}^l_k)^T\bm{h}^m\big)$. The FN $f_{y(l)}^{m,k}$ represents the pdf of the received signal ${y}^m_k(l)$, i.e. $f_{y(l)}^{m,k}(z^{m,k}_l,\lambda_k)=\mathcal{CN}\big({y}^m_k(l);z^{m,k}_l,\lambda_k^{-1}\big)$. In addition, since the noise samples between different antennas are also independent, each noise precision VN $\lambda_k$ is also connected to the FNs $\big\{f^{m^\prime,k}_{y(l)}, l\in\mathcal{L}_p\big\}$ located on the PUD factor graph of the $m^\prime$-th antenna, where $m^\prime\in\mathcal{M}/m$.  
\begin{figure}
	\centering
	\includegraphics[width=1\columnwidth]{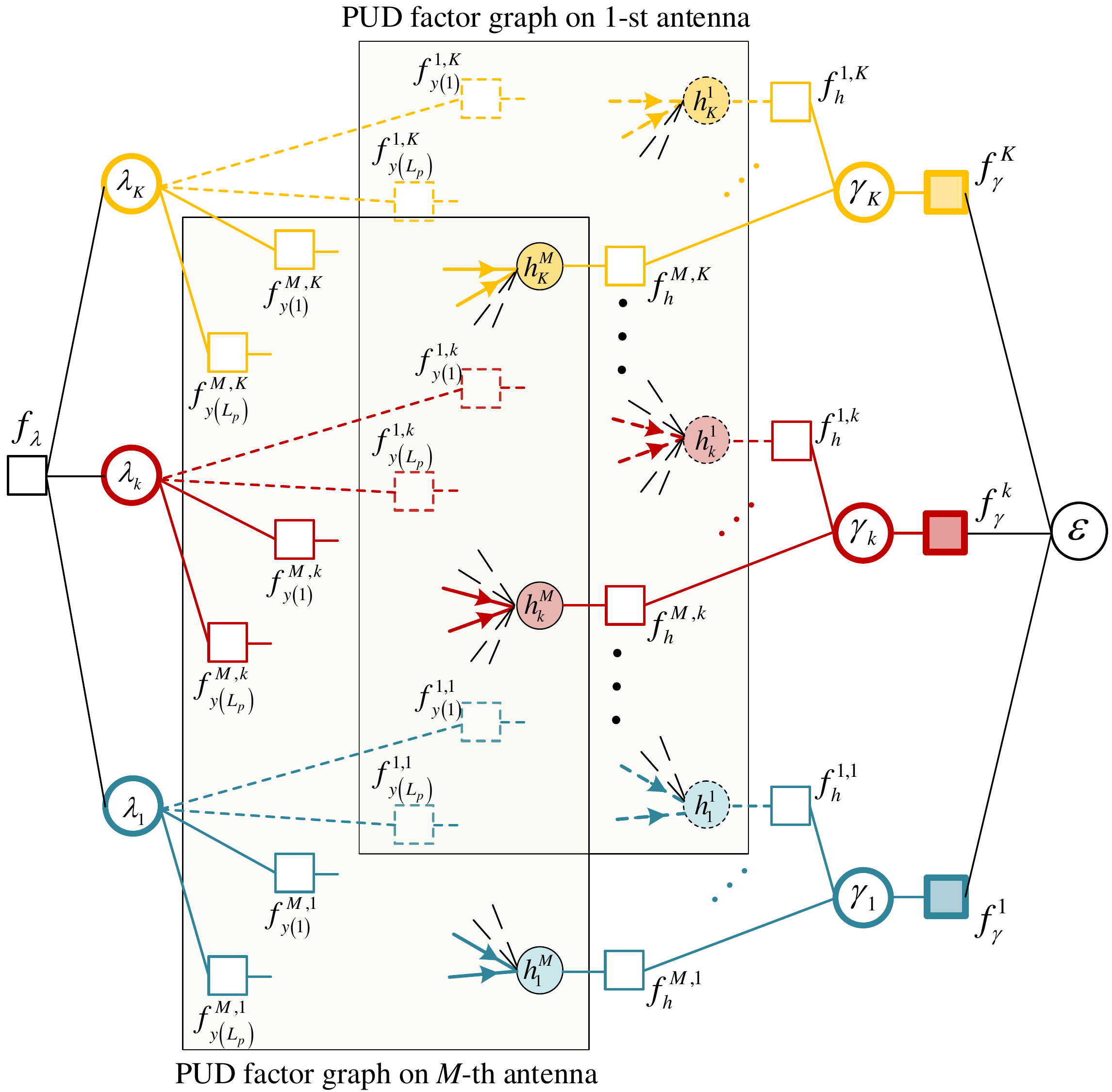}
	\caption{Factor graph representation of the posterior probability factorization.}
	\label{factorgraph}
\end{figure}

The principle of partially uni-directional message passing is explained as follows. As shown in Fig. \ref{PUD}, the VNs $\{h^m_k,k\in\mathcal{K}\}$ and FNs $\{f^{m,k}_{\delta(l)},l\in\mathcal{L}_p, k\in\mathcal{K}\}$ are fully connected. However, only the messages passed from the FNs $\{f^{m,k}_{\delta(l)}, l\in\mathcal{L}_p\}$ are employed to update the belief of $h^m_k$. In other words, only the messages obtained from the $k$-th matched filter are used to update the channel estimate of UE $k$. On the other hand, the messages passed along the dashed edges in Fig. \ref{PUD} are not used by UE $k$ to update $h^m_k$, since these messages are obtained from the other matched filters $\{k^\prime\in\mathcal{K}/k\}$, and therefore correlated with the messages from the $k$-th matched filter. Apart from the messages passed between FNs $\{f^{m,k}_{\delta(l)},l\in\mathcal{L}_p, k\in\mathcal{K}\}$ and VNs $\{h^m_k,k\in\mathcal{K}\}$, the remaining message passing procedures between the VNs and FNs in Fig. \ref{PUD} are still bi-directional, which explains why this factor graph is \emph{partially} uni-directional. The detailed message passing procedure will be explained in Section \ref{FMP}. 

Based on the PUD factor graph in Fig. \ref{PUD}, we establish a complete factor graph in Fig. \ref{factorgraph} for the factorization in equation ($d$) of (\ref{APP}). This factor graph is composed of $M$ parallel PUD factor graphs, and additional sets of VNs $\{\gamma_k,k\in\mathcal{K}\}$, $\epsilon$, and FNs $\{f_h^{m,k},m\in\mathcal{M},k\in\mathcal{K}\}$, $\{f_\gamma^k,k\in\mathcal{K}\}$, $f_\lambda$. Here, $\{\gamma_k,k\in\mathcal{K}\}$ and $\epsilon$ are related to the prior variance of the VNs $\{h^m_k,k\in\mathcal{K},M\in\mathcal{M}\}$. FN $f_h^{m,k}$ and $f_\lambda$ represent the pdf $f_h^{m,k}(h^m_k,\gamma_k)=p(h^m_k|\gamma_k)$ in (\ref{APP}) and $p(\lambda)\propto 1/\lambda$, while each FN $f_\gamma^k$ represents the Gamma-distribution pdf $f_\gamma^k(\gamma_k,\epsilon)=p(\gamma_k|\epsilon)=Gam(\gamma_k;\epsilon,\eta)$. For reading convenience, FNs and related functions on the factor graph are summarized in TABLE \ref{FNMeaning}.

\begin{remark}\label{rmk}
The pdf $p(\gamma_k)$ in equation ($d$) of (\ref{APP}) is represented with a FN $f_\gamma^k$ in the factor graph, while this FN $f_\gamma^k$ is connected to a learnable shape parameter VN $\epsilon$. In the message passing procedure described below, the shape parameter $\epsilon$ is updated iteratively. According to \cite{Justify_epsilon}, this shape parameter $\epsilon$ functions as a selective amplifier for the VNs $\{\gamma_k,k\in\mathcal{K}\}$. Therefore, updating the shape parameter VN $\epsilon$ benefits the robustness of the proposed PUDMP-SBL-aUADCE algorithm, especially under the case with non-ideal delay information. Related details will be explained in Section \ref{nonidealTA}.
\end{remark}
\begin{table}
\renewcommand\arraystretch{1.3}
\caption{FNs in the factor graph and related functions}
\centering
\begin{tabular}{ccc}
	\Xhline{1.2pt}
	FN&Function&PDF\\
	\hline
	$f_{\lambda}$&$p(\lambda)$&$\propto \lambda^{-1}$\\
	$f^{m,k}_{y(l)}$&$f_{y(l)}^{m,k}(z^{m,k}_l,\lambda_k)$&$\mathcal{CN}\big({y}^m_k(l);z^{m,k}_l,{\lambda^{-1}_k}\big)$\\
	$f^{m,k}_{\delta(l)}$&$f^{m,k}_{\delta(l)}(z^{m,k}_l,\bm{h}^m)$&$\delta\big(z^{m,k}_l-(\bar{\bm{p}}^l_k)^T\bm{h}^m\big)$\\
	$f_h^{m,k}$&$p(h^m_k|\gamma_k)$&$\mathcal{CN}(h^m_k;0,\gamma^{-1}_k)$\\
	$f_\gamma^k$&$p(\gamma_k|\epsilon)$&$Gam(\gamma_k;\epsilon,\eta)$\\
	\Xhline{1.2pt}
\end{tabular}
\label{FNMeaning}
\end{table}
\subsection{Message Passing Procedure and Message Update Rules}\label{FMP}
In this subsection, we explain the \emph{forward} (from the left to the right) message passing procedure and the \emph{backward} (from the right to the left) message passing procedure in Fig. \ref{factorgraph}, as well as related message update rules at different VNs and FNs. 
\subsubsection{Messages Update at FN $f^{m,k}_{\delta(l)}$} We first consider the forward message passing procedure at FN $f^{m,k}_{\delta(l)}$ in the $t$-th iteration, and the forward message $I^t_{f^{m,k}_{\delta(l)}\to h^m_k}$ passed from FN $f^{m,k}_{\delta(l)}$ to VN $h^m_k$ is derived as
\begin{equation}\label{VN_delta_F}
\begin{split}
I^t_{f^{m,k}_{\delta(l)}\to h^m_k}(h^m_k)&\!=\!\int f^{m,k}_{\delta(l)}(z^{m,k}_l,\bm{h}^m)I^{t}_{z^{m,k}_l\to f^{m,k}_{\delta(l)}}(z^{m,k}_l)\\
&\!\times\!\!\prod\limits_{k^\prime\neq k}\!\!I^{t-1}_{h^m_{k^\prime}\to f^{m,k}_{\delta(l)}}(h^m_{k^\prime})dz^{m,k}_ld\{h^m_{k^\prime},k^\prime\in\mathcal{K}/k\},\\
&\!=\!\mathcal{CN}(h^m_k;\mu^t_{f^{m,k}_{\delta(l)}\to h^m_k},v^t_{f^{m,k}_{\delta(l)}\to h^m_k}),
\end{split}
\end{equation}
where 
\begin{equation}\label{VN_delta_F_muv}
\begin{split}
v^t_{f^{m,k}_{\delta(l)}\to h^m_k}&=\Big({\hat{\lambda}_k}^{-1}+\sum\limits_{k^\prime\neq k}|\bar{p}^l_{k,k^\prime}|^2v^{t-1}_{h^m_{k^\prime}\to f^{m,k}_{\delta(l)}}\Big)\Big/{|\bar{p}^l_{k,k}|^2},\\
\mu^t_{f^{m,k}_{\delta(l)}\to h^m_k}&=\Big(y^m_k(l)-\sum\limits_{k^\prime\neq k}\bar{p}^l_{k,k^\prime}\mu^{t-1}_{h^m_{k^\prime}\to f^{m,k}_{\delta(l)}}\Big)\Big/{\bar{p}^l_{k,k}},
\end{split}
\end{equation}
where $I^{t}_{z^{m,k}_l\to f^{m,k}_{\delta(l)}}(z^{m,k}_l)=\mathcal{CN}(z^{m,k}_l;y^m_k(l),{\hat{\lambda}_k}^{-1})$ is the forward message passed from VN $z^{m,k}_l$ to FN $f^{m,k}_{\delta(l)}$, which will be later derived in (\ref{Z_F}). $I^{t-1}_{h^m_{k^\prime}\to f^{m,k}_{\delta(l)}}(h^m_{k^\prime})=\mathcal{CN}(h^m_{k^\prime};\mu^{t-1}_{h^m_{k^\prime}\to f^{m,k}_{\delta(l)}},v^{t-1}_{h^m_{k^\prime}\to f^{m,k}_{\delta(l)}})$ is the backward message passed from VN $h^m_{k^\prime}$ to FN $f^{m,k}_{\delta(l)}$ in the $(t-1)$-th iteration, which will be later derived in (\ref{VN_h_B3}). 

Similarly, we can also derive the backward message $I^t_{f^{m,k}_{\delta(l)}\to z^{m,k}_l}(z^{m,k}_l)$ passed from FN $f^{m,k}_{\delta(l)}$ to VN $z^{m,k}_l$ in the $t$-th iteration, i.e.
\begin{equation}\label{delta2z}
\begin{split}
&I^t_{f^{m,k}_{\delta(l)}\to z^{m,k}_l}(z^{m,k}_l)\\
&=\int f^{m,k}_{\delta(l)}(z^{m,k}_l,\bm{h}^m)\prod\limits_{k^\prime\in\mathcal{K} }I^{t}_{h^m_{k^\prime}\to f^{m,k}_{\delta(l)}}(h^m_{k})d\{h^m_{k},k\in \mathcal{K}\},\\
&=\mathcal{CN}(z^{m,k}_l;\mu^t_{f^{m,k}_{\delta(l)}\to z^{m,k}_l},v^t_{f^{m,k}_{\delta(l)}\to z^{m,k}_l}),
\end{split}
\end{equation}
where
\begin{equation}\label{delta2z_muv}
\begin{split}
v^t_{f^{m,k}_{\delta(l)}\to z^{m,k}_l}&=\sum\limits_{k^\prime\in\mathcal{K}}|\bar{p}^l_{k,k^\prime}|^2v^{t}_{h^m_{k^\prime}\to f^{m,k}_{\delta(l)}},\\
\mu^t_{f^{m,k}_{\delta(l)}\to z^{m,k}_l}&=\sum\limits_{k^\prime\in\mathcal{K}}\bar{p}^l_{k,k^\prime}\mu^{t}_{h^m_{k^\prime}\to f^{m,k}_{\delta(l)}}.
\end{split}
\end{equation}
\subsubsection{Messages Update at VN $h^{m}_k$} For each VN $h^m_k$, its belief $b^t(h^{m}_k)$ in the $t$-th iteration is obtained as
\begin{equation}\label{belief}
\begin{split}
b^t(h^{m}_k)&\propto I^t_{f^{m,k}_{h}\to h^{m}_k}(h^{m}_k)\prod_{l\in\mathcal{L}_p}I^t_{f^{m,k}_{\delta(l)}\to h^{m}_k}(h^{m}_k),\\
&\propto \mathcal{CN}(h^{m}_k;\mu^t_{h^{m}_k},v^t_{h^{m}_k}),
\end{split}
\end{equation}
with
\begin{equation}\label{belief_muv}
\begin{split}
v^t_{h^{m}_k}&=\Big(\sum\limits_{l\in\mathcal{L}_p}\frac{1}{v^t_{f^{m,k}_{\delta(l)}\to h^m_k}} + \hat{\gamma}_k\Big)^{-1},\\
\mu^t_{h^{m}_k}&=v^t_{h^{m}_k}\Bigg(\sum\limits_{l\in\mathcal{L}_p}\frac{\mu^t_{f^{m,k}_{\delta(l)}\to h^m_k}}{v^t_{f^{m,k}_{\delta(l)}\to h^m_k}}\Bigg),
\end{split}
\end{equation}
where $I^t_{f^{m,k}_{h}\to h^{m}_k}(h^{m}_k)=\mathcal{CN}(h^{m}_k;0,\hat{\gamma}^{-1}_k)$ is the message passed from FN $f^{m,k}_{h}$ to VN $h^{m}_k$, which will be derived in (\ref{FH_B}). The term $\prod_{l\in\mathcal{L}_p}I^t_{f^{m,k}_{\delta(l)}\to h^{m}_k}(h^{m}_k)$ implies that the belief of $h^{m}_k$ is updated according to the principle of \emph{partially uni-directional} message passing in Fig. \ref{PUD}. Then, we derive the following backward messages passed from VN $h^{m}_k$ to different FNs $f^{m,k}_{\delta(l)}$ and $f^{m,k^\prime}_{\delta(l)}$. 

Denote $I^t_{h^{m}_k\to f^{m,k}_{\delta(l)}}(h^{m}_k)$ as the backward message passed from the VN $h^{m}_k$ of the $k$-th UE to FN $f^{m,k}_{\delta(l)}$ in the $k$-th matched filter, and it is derived as 
\begin{equation}\label{VN_h_B1}
I^t_{h^{m}_k\to f^{m,k}_{\delta(l)}}(h^{m}_k)\propto \frac{b^t(h^{m}_k)}{I^{t}_{f^{m,k}_{\delta(l)}\to h^m_k}(h^m_k)}.
\end{equation}

Denote $I^t_{h^{m}_k\to f^{m,k^\prime}_{\delta(l)}}(h^{m}_k)$ as the backward message passed from the VN $h^{m}_k$ of the $k$-th UE to the FN $f^{m,k^\prime}_{\delta(l)}$ in the $k^\prime$-th matched filter, and it is derived as 
\begin{equation}\label{VN_h_B2}
I^t_{h^{m}_k\to f^{m,k^\prime}_{\delta(l)}}(h^{m}_k)\!\propto\! \left\{ \begin{array}{l}
\! \! \frac{b^t(h^{m}_k)}{I^{t}_{f^{m,k}_{\delta(l-1)}\! \to\!  h^m_k}\!\!  (h^m_k)I^{t}_{f^{m,k}_{\delta(l)}\! \to\!  h^m_k}\!\!  (h^m_k)},\ \text{for }k^\prime<k,\\
\! \! \frac{b^t(h^{m}_k)}{I^{t}_{f^{m,k}_{\delta(l)}\! \to\! h^m_k}\!\! (h^m_k)I^{t}_{f^{m,k}_{\delta(l+1)}\! \to\! h^m_k}\!\! (h^m_k)},\ \text{for }k^\prime>k.
\end{array} \right.
\end{equation}
We take the case with $k^\prime<k$ to explain the derivation in (\ref{VN_h_B2}). According to (\ref{VN_delta_F_muv}), the forward messages $I^{t}_{f^{m,k}_{\delta(l-1)}\to h^m_k}(h^m_k)$ and $I^{t}_{f^{m,k}_{\delta(l)}\to h^m_k}(h^m_k)$ involve the sampled output $y^m_{k}(l-1)$ and $y^m_{k}(l)$, respectively. It is shown in Fig. \ref{MUI} that both $y^m_{k}(l-1)$ and $y^m_{k}(l)$ are correlated with $y^m_{k^\prime}(l)$ for $k^\prime<k$. Therefore, the forward messages $I^{t}_{f^{m,k}_{\delta(l-1)}\to h^m_k}(h^m_k)$ and $I^{t}_{f^{m,k}_{\delta(l)}\to h^m_k}(h^m_k)$ are also correlated with $y^m_{k^\prime}(l)$. When $y^m_{k^\prime}(l)$ is used to calculate $I^{t+1}_{f^{m,k^\prime}_{\delta(l)}\to h^m_{k^\prime}}(h^m_{k^\prime})$ at FN $ f^{m,k^\prime}_{\delta(l)}$ as in (\ref{VN_delta_F}), the incoming message $I^t_{h^{m}_k\to f^{m,k^\prime}_{\delta(l)}}(h^{m}_k)$ should be independent from $y^m_{k^\prime}(l)$ to avoid the correlation problem. Therefore, $I^{t}_{f^{m,k}_{\delta(l-1)}\to h^m_k}(h^m_k)$ and $I^{t}_{f^{m,k}_{\delta(l)}\to h^m_k}(h^m_k)$ should be excluded from $b^t(h_k^m)$ to obtain the backward message $I^t_{h^{m}_k\to f^{m,k^\prime}_{\delta(l)}}(h^{m}_k)$ in (\ref{VN_h_B2}). Similarly, $I^{t}_{f^{m,k}_{\delta(l)}\to h^m_k}(h^m_k)$ and $I^{t}_{f^{m,k}_{\delta(l+1)}\to h^m_k}(h^m_k)$ should be excluded from $b^t(h_k^m)$ for $k^\prime> k$. According to (\ref{VN_h_B1}) and (\ref{VN_h_B2}), the backward message update at VN $h^{m}_k$ is tedious for different UE indexes. Sacrificing a little useful information, we obtain a more unified message update rule for $\forall k^\prime\in\mathcal{K}$, i.e.,
\begin{equation}\label{VN_h_B3}
\begin{split}
I^t_{h^{m}_k\to f^{m,k^\prime}_{\delta(l)}}(h^{m}_k)&\propto \frac{b^t(h^{m}_k)}{\prod\limits_{\zeta=l-1}^{l+1}I^{t}_{f^{m,k}_{\delta(\zeta)}\to h^m_k}(h^m_k)},\\
&=\mathcal{CN}(h^{m}_k;\mu^t_{h^{m}_k\to f^{m,k^\prime}_{\delta(l)}},v^t_{h^{m}_k\to f^{m,k^\prime}_{\delta(l)}}),
\end{split}
\end{equation}
with
\begin{equation}\label{VN_h_B3_muv}
\begin{split}
v^t_{h^{m}_k\to f^{m,k^\prime}_{\delta(l)}}&=\Big(\frac{1}{v^t_{h^{m}_k}} - \sum\limits_{\zeta=l-1}^{l+1} \frac{1}{v^{t}_{f^{m,k}_{\delta(\zeta)}\to h^m_k}}   \Big)^{-1},\\
\mu^t_{h^{m}_k\to f^{m,k^\prime}_{\delta(l)}}&=v^t_{h^{m}_k}\Bigg(\frac{\mu^t_{h^{m}_k}}{v^t_{h^{m}_k}} - \sum\limits_{\zeta=l-1}^{l+1} \frac{\mu^{t}_{f^{m,k}_{\delta(\zeta)}\to h^m_k}}{v^{t}_{f^{m,k}_{\delta(\zeta)}\to h^m_k}}\Bigg).
\end{split}
\end{equation}
In other words, we exclude three messages $I^{t}_{f^{m,k}_{\delta(l-1)}\to h^m_k}(h^m_k)$, $I^{t}_{f^{m,k}_{\delta(l)}\to h^m_k}(h^m_k)$, and $I^{t}_{f^{m,k}_{\delta(l+1)}\to h^m_k}(h^m_k)$ from the belief $b^t(h^{m}_k)$ to obtain the backward message from VN $h^m_k$ to an arbitrary FN $f^{m,k^\prime}_{\delta(l)}$ with $\forall k^\prime \in\mathcal{K}$. Replacing (\ref{VN_h_B1}) and (\ref{VN_h_B2}) with (\ref{VN_h_B3}) may lose some useful uncorrelated information, while the loss is negligible if $L_p\gg 1$. 
\subsubsection{Message Update at FN $f_h^{m,k}$, VN $\gamma_k$, FN $f_\gamma^k$, and VN $\epsilon$} The forward message $I^t_{f_h^{m,k}\to \gamma_k}(\gamma_k)$ passed from FN $f_h^{m,k}$ to VN $\gamma_k$ is derived according to the mean-field (MF) rule, i.e.
\begin{equation}\label{FN_fh_F}
\begin{split}
I^t_{f_h^{m,k}\to \gamma_k}(\gamma_k)&=\exp \Bigg\{\int b^t(h^{m}_k)\ln f_h^{m,k}(h^m_k,\gamma_k)d h^{m}_k\Bigg\},\\
&\propto \gamma_k\exp \Big\{-\gamma_k\Big(|\mu^t_{h^{m}_k}|^2+v^t_{h^{m}_k}\Big)\Big\}.
\end{split}
\end{equation}
The belief $b^t(\gamma_k)$ of $\gamma_k$ is obtained by combining all the incoming messages to VN $\gamma_k$,
\begin{equation}\label{b_gamma}
\begin{split}
b^t(\gamma_k)&\propto I^{t-1}_{f^k_\gamma\to \gamma_k}(\gamma_k)\prod\limits^M_{m=1}I^t_{f_h^{m,k}\to \gamma_k}(\gamma_k),\\
&\propto {\gamma_k}^{\hat{\epsilon}+M-1}\exp\Bigg\{\!-\!\gamma_k\Bigg[\eta+\sum\limits_{m=1}^M\Big(|\mu^t_{h^{m}_k}|^2\!+\!v^t_{h^{m}_k}\Big)\Bigg]\Bigg\},
\end{split}
\end{equation}
where $I^{t-1}_{f^k_\gamma\to \gamma_k}(\gamma_k)=Gam(\gamma_k;\hat{\epsilon},\eta)$ is the backward message from FN $f_\gamma^k$ to VN $\gamma_k$ in the previous iteration, which will be derived in (\ref{FN_gamma_B}). Then, the hyper-parameter $\hat{\gamma}_k$ is estimated as
\begin{equation}\label{Gam_up}
\hat{\gamma}_k\overset{\Delta}{=}\mathbb{E}_{b^t(\gamma_k)}\big[\gamma_k\big]=\frac{\hat{\epsilon}+M}{\eta+\sum\limits_{m=1}^M\Big(|\mu^t_{h^{m}_k}|^2+v^t_{h^{m}_k}\Big)},
\end{equation}
and $\eta$ is commonly set as $0$ for the SBL method \cite{Justify_epsilon}

With the belief $b^t(\gamma_k)$ of $\gamma_k$, we can obtain the backward message $I^t_{f_h^{m,k}\to h^m_k}(h^m_k)$ passed from FN $f_h^{m,k}$ to VN $h^m_k$, which is derived according to the following MF rule,
\begin{equation}\label{FH_B}
\begin{split}
I^t_{f_h^{m,k}\to h^m_k}(h^m_k)&=\exp\Bigg\{\int b^t(\gamma_k)\ln f_h^{m,k}(h^m_k,\gamma_k)d \gamma_k\Bigg\},\\
&\propto \mathcal{CN}(h^m_k;0,\hat{\gamma}^{-1}_k).
\end{split}
\end{equation}

We further employ the MF rule to derive the forward message $I^t_{f_\gamma^k\to \epsilon}(\epsilon)$ passed from FN $f_\gamma^k$ to VN $\epsilon$, i.e.
\begin{equation}
\begin{split}
I^t_{f_\gamma^k\to \epsilon}(\epsilon)&=\exp \Bigg\{\int b^t(\gamma_k)\ln f_\gamma^k(\gamma_k,\epsilon) d\gamma_k\Bigg\},\\
&\propto \frac{\eta^\epsilon}{\Gamma(\epsilon)}\exp\Bigg\{(\epsilon-1)\mathbb{E}_{b^t(\gamma_k)}\big[\ln\gamma_k\big]-\eta\hat{\gamma}_k\Bigg\}.
\end{split}
\end{equation}
Combine all incoming messages to VN $\epsilon$, the belief $b^t(\epsilon)$ is 
\begin{equation}
\begin{split}
b^t(\epsilon)&\propto\prod\limits_{k=1}^{K}I^t_{f_\gamma^k\to \epsilon}(\epsilon),\\
&\propto\Big(\frac{\eta^\epsilon}{\Gamma(\epsilon)}\Big)^K\exp\Bigg\{(\epsilon\!-\!1)\sum\limits_{k=1}^K\Big(\mathbb{E}_{b^t(\gamma_k)}\big[\ln\gamma_k\big]\!-\!\eta\hat{\gamma}_k\Big)\Bigg\},
\end{split}
\end{equation}
and the backward message $I^t_{f_\gamma^k\to \gamma_k}(\gamma_k)$ passed from FN $f_\gamma^k$ to VN $\gamma_k$ is further derived as
\begin{equation}\label{FN_gamma_B}
\begin{split}
I^t_{f_\gamma^k\to \gamma_k}(\gamma_k)&=\exp \Bigg\{\int b^t(\epsilon)\ln f_\gamma^k(\gamma_k,\epsilon) d\epsilon\Bigg\},\\
&\propto Gam(\gamma_k;\hat{\epsilon},\eta),
\end{split}
\end{equation}
where $\hat{\epsilon}\overset{\Delta}{=}\mathbb{E}_{b^t(\epsilon)}\big[\epsilon\big]$. However,  it is difficult to directly calculate $\hat{\epsilon}$ by definition. Here, we use another empirical yet effective solution \cite{Justify_epsilon} to update $\hat{\epsilon}$ in a more straightforward way, i.e.,
\begin{equation}\label{eps_up}
\hat{\epsilon}=\frac{1}{2}\sqrt{\ln\Bigg(\frac{1}{K}\sum\limits_{k=1}^{K}\hat{\gamma}_k\Bigg)-\frac{1}{K}\sum\limits_{k=1}^{K}\ln\hat{\gamma}_k}.
\end{equation}  
\subsubsection{Message Update at FN $f^{m,k}_{y(l)}$ and VN $z^{m,k}_l$} The backward message $I^t_{f^{m,k}_{y(l)}\to z^{m,k}_l}$ passed from FN $f^{m,k}_{y(l)}$ to VN $z^{m,k}_l$ is derived as
\begin{equation}
\begin{split}
I^t_{f^{m,k}_{y(l)}\to z^{m,k}_l}(z^{m,k}_l)&\!=\!\exp\Bigg\{\!\int b^t\big(\lambda_k\big)\ln f_{y(l)}^{m,k}\big(z^{m,k}_l,\lambda_k\big) d \lambda_k\!\Bigg\},\\
&\propto \mathcal{CN}\Big(z^{m,k}_l;y^m_k(l),\hat{\lambda}^{-1}_k\Big),
\end{split}
\end{equation}
where $b^t(\lambda_k)$ and $\hat{\lambda}^{-1}_k$ will be later explained in (\ref{B_Lambda}) and (\ref{E_Lambda}), respectively. 

For each VN $z^{m,k}_l$, the forward message $I^t_{z^{m,k}_l \to f^{m,k}_{\delta(l)}} (z^{m,k}_l)$ is simply obtained as
\begin{equation}\label{Z_F}
I^t_{z^{m,k}_l \to f^{m,k}_{\delta(l)}} (z^{m,k}_l)=I^t_{f^{m,k}_{y(l)}\to z^{m,k}_l}(z^{m,k}_l).
\end{equation}
In this way, the belief $b^t(z^{m,k}_l)$ of $z^{m,k}_l$ is obtained as
\begin{equation}\label{beliefZ}
\begin{split}
b^t(z^{m,k}_l)&\propto I^t_{f^{m,k}_{y(l)}\to z^{m,k}_l}(z^{m,k}_l)I^t_{f^{m,k}_{\delta(l)} \to z^{m,k}_l} (z^{m,k}_l),\\
&\propto\mathcal{CN}(z^{m,k}_l;\mu^t_{z^{m,k}_l},v^t_{z^{m,k}_l}),
\end{split}
\end{equation}
with
\begin{equation}\label{beliefZ_muv}
\begin{split}
v^t_{z^{m,k}_l}&= \Bigg(\frac{1}{v^t_{f^{m,k}_{\delta(l)}\to z^{m,k}_l}} + \hat{\lambda}_k\Bigg)^{-1},\\
\mu^t_{z^{m,k}_l}&=v^t_{z^{m,k}_l}\Bigg( y^m_k(l)\hat{\lambda}_k + \frac{\mu^t_{f^{m,k}_{\delta(l)}\to z^{m,k}_l}}{v^t_{f^{m,k}_{\delta(l)}\to z^{m,k}_l}} \Bigg).
\end{split}
\end{equation}

With the belief $b^t(z^{m,k}_l)$, we can now calculate the forward message $I^t_{f^{m,k}_{y(l)}\to \lambda_k}(\lambda_k)$ passed from FN $f^{m,k}_{y(l)}$ to VN $\lambda_k$. According to the MF rule, we have
\begin{equation}
\begin{split}
I^t_{f^{m,k}_{y(l)}\to \lambda_k}(\lambda_k)&\!=\!\exp\Bigg\{\!\int b^t\big(z^{m,k}_l\big)\ln f_{y(l)}^{m,k}\big(z^{m,k}_l,\lambda_k\big) d z^{m,k}_l\!\Bigg\},\\
&\propto \mathcal{CN}\Big(z^{m,k}_l;y^m_k(l),\hat{\lambda}^{-1}_k\Big).
\end{split}
\end{equation}
Then, the belief $b^t(\lambda_k)$ of each VN $\lambda_k$ is derived as
\begin{equation}
\begin{split}\label{B_Lambda}
&b^t(\lambda_k)\propto f(\lambda)\prod\limits^M_{m=1}\prod\limits^{L_p}_{l=1}I^t_{f^{m,k}_{y(l)}\to \lambda_k}(\lambda_k),\\
&\propto \lambda_k^{ML_p-1}\exp\!\Bigg\{\!\!-\!\lambda_k\!\sum\limits^M_{m=1}\sum\limits^{L_p}_{l=1}\Big[\big|y^m_k(l)\!-\!\mu^t_{z^{m,k}_l}\big|^2+v^t_{z^{m,k}_l}\Big]\Bigg\}.
\end{split}
\end{equation}
In this way, the noise precision $\hat{\lambda}_k$ is estimated as
\begin{equation}\label{E_Lambda}
\hat{\lambda}_k\overset{\Delta}{=}\mathbb{E}_{b^t(\lambda_k)}\Big[\lambda_k\Big]=\frac{ML_p}{\sum\limits^M_{m=1}\sum\limits^{L_p}_{l=1}\Big[\big|y^m_k(l)\!-\!\mu^t_{z^{m,k}_l}\big|^2+v^t_{z^{m,k}_l}\Big]}.
\end{equation}
\subsection{Algorithm Summary and Complexity Analysis}\label{complex}
\begin{algorithm}[t!]\setstretch{1.0}
	
	\caption{PUDMP-SBL-aUADCE Algorithm}
	
	\label{alg:MPaSBL}
	
	
	\KwIn{$\{\bm{Y}^m,m\in\mathcal{M}\}$, $\{\bm{\bar{P}}_k,k\in\mathcal{K}\}$}
	
	
	\KwOut{UAD $\{\hat{\alpha}_{k},k\in\mathcal{K}\}$ and CE $\{\hat{h}^m_k,m\in\mathcal{M},k\in\mathcal{K}\}$}
	
	{\textbf{Initialize:}} 
	
	{\ \; $\text{Set}\ \hat{\epsilon}=10^{-3},\ \hat{\lambda}_k=1,\ \hat{\gamma}_k=1\ \text{for}\ \forall k\text{, and set} \ t=0.$ }
	
	{\ \ \ \ $\text{Set} \ v^t_{h^{m}_k\to f^{m,k^\prime}_{\delta(l)}}=1,\ \mu^t_{h^{m}_k\to f^{m,k^\prime}_{\delta(l)}}=0 \ \text{for}\ \forall k,\ \forall k^\prime, \ \forall l,\ \forall m$.}

	\For{$t=1:N_{it}$}{
		
		1. Update $I^t_{f^{m,k}_{\delta(l)}\to h^m_k}(h^m_k)$ with (\ref{VN_delta_F}) and (\ref{VN_delta_F_muv}).
		
		2. Update $b^t(h^{m}_k)$ with (\ref{belief}) and (\ref{belief_muv}).
		
		3. Update $\hat{\gamma}_k$ with (\ref{Gam_up}).
		
		4. Update $\hat{\epsilon}$ with (\ref{eps_up}).
		
		5. Update $I^t_{h^{m}_k\to f^{m,k^\prime}_{\delta(l)}}(h^{m}_k)$ with (\ref{VN_h_B3}) and (\ref{VN_h_B3_muv}).
		
		6. Update $I^t_{f^{m,k}_{\delta(l)}\to z^{m,k}_l}(z^{m,k}_l)$ with (\ref{delta2z}) and (\ref{delta2z_muv}).
		
		7. Update $b^t(z^{m,k}_l)$ with (\ref{beliefZ}) and (\ref{beliefZ_muv}).
		
		8. Update $\hat{\lambda}_k$ with (\ref{E_Lambda}).
		
	}
	\textbf{UAD and CE Decision:} If $\hat{\gamma}_k<\gamma_{th}$, $\hat{\alpha}_k=1$ and $\hat{h}^m_k=\mu^{N_{it}}_{h^{m}_k}$ for $\forall m\in\mathcal{M}$. Otherwise, $\hat{\alpha}_k=0$ and $\hat{h}^m_k=0$ for $\forall m\in\mathcal{M}$. 
	
	
\end{algorithm}
According to the descriptions above, the proposed PUDMP-SBL-aUADCE algorithm is summarized as in Algorithm \ref{alg:MPaSBL}. The input information includes the received pilot signal $\bm{Y}^m$ on all the $M$ antennas, and the equivalent pilot matrix $\bm{\bar{P}}_k$ of each matched filter $k\in\mathcal{K}$. Here, the element $\bar{p}^l_{k,k^\prime}$ on the $l$-th row and $k^\prime$-th column of matrix $\bar{\bm{P}}_k$ is given in (\ref{eff_pilot}). After the initialization step, the message passing procedure of the PUDMP-SBL-aUADCE algorithm is performed with $N_{it}$ iterations, producing the activity-detection result $\{\hat{\alpha}_k,k\in\mathcal{K}\}$, where $\hat{\alpha}_k=1$ if $\hat{\gamma}_k<\gamma_{th}$, and $\hat{\alpha}_k=0$ otherwise. The UAD threshold is empirically set as $\gamma_{th}=10$ for the simulations in Section \ref{simu_section}. We further obtain the CE result for the UEs that are detected as active, i.e. $\hat{h}^m_k=\mu^{N_{it}}_{h^{m}_k}$ if $\hat{\alpha}_k=1$, where $\mu^{N_{it}}_{h^{m}_k}$ is the mean of the belief distribution $b^t(h^m_k)$ in the $N_{it}$-th iteration, which is given by (\ref{belief_muv}). 

We further analyze the computational complexity of the proposed PUDMP-SBL-aUADCE algorithm, which is assessed by the number of multiplication/division operations. According to the partially uni-directional message passing principle, the message $I^t_{f^{m,k}_{\delta(l)}\to h^m_k}(h^m_k)$ in step 1 of Algorithm \ref{alg:MPaSBL} is independently updated for different UE $k$. Considering each UE index $k$, pilot-symbol index $l$, and antenna index $m$, updating $I^t_{f^{m,k}_{\delta(l)}\to h^m_k}(h^m_k)$ totally incurs $\mathcal{O}(K^2L_pM)$ operations in each iteration. Updating $b^t(h^{m}_k)$ and $\hat{\gamma}_k$ in step 2 and step 3 incurs $\mathcal{O}(KL_pM)$ operations and $\mathcal{O}(KM)$ operations, respectively. According to (\ref{eps_up}), the update of $\hat{\epsilon}$ in step 4 only needs $K+1$ logarithmic operations and $1$ additional operation to take the square root. Then, $\mathcal{O}(KL_pM)$ operations are required to update $I^t_{h^{m}_k\to f^{m,k^\prime}_{\delta(l)}}(h^{m}_k)$ in step 5, while $\mathcal{O}(K^2L_pM)$ operations are required for $I^t_{f^{m,k}_{\delta(l)}\to z^{m,k}_l}(z^{m,k}_l)$ in step 6. Finally, updating both $b^t(z^{m,k}_l)$ in step 7 and $\hat{\lambda}_k$ in step 8 will incur $\mathcal{O}(KL_pM)$ operations. To sum up, the proposed PUDMP-SBL-aUADCE algorithm totally incurs $\mathcal{O}(N_{it}K^2L_pM)$ multiplication/division operations. 

Despite that the overall computational complexity scales with the square of the UE number $K$, it is noted that the computationally-intensive messages $I^t_{f^{m,k}_{\delta(l)}\to h^m_k}(h^m_k)$ and $I^t_{f^{m,k}_{\delta(l)}\to z^{m,k}_l}(z^{m,k}_l)$ are \emph{independently} updated for different UE $k$, which facilitates parallel implementation. Therefore, we can allocate the computation task to $K$ distributed processors, which is one major advantage of the proposed message passing algorithm. In this way, the computational complexity at each local processor still scales \emph{linearly} with the UE number $K$, the pilot length $L_p$, and the antenna number $M$.
\subsection{SINR Analysis for Synchronous and Asynchronous Case}\label{SINRana}
The synchronous case is equivalent to setting the same delay $\tau_k=\tau$ for $\forall k\in\mathcal{K}$. In this case, the proposed PUDMP-SBL-aUADCE algorithm  will reduce to the MP-BSBL algorithm in \cite{ZYYMP}. To make this paper self-contained, we explain how to modify the PUDMP-SBL-aUADCE algorithm in Algorithm \ref{alg:MPaSBL} for synchronous GF-RA. Firstly, we adopt the same matched filter to all the UEs, so that the received signal in (\ref{rcv_vec}) will be re-written as $\bm{y}^m_\text{syn}=\bm{P}\bm{h}^m+\bm{n}^m$, where $\bm{P}$ is the pilot matrix with element $p^l_k$ on its $l$-th row and $k$-th column. In this way, the factor graph in Fig. \ref{PUD} will contain $K$ VNs $\{h^m_k,k\in\mathcal{K}\}$, and only $L_p$ FNs $\{f^m_{\delta(l)},l\in\mathcal{L}_p\}$, since we have $f^{m,k}_{\delta(l)}=f^m_{\delta(l)}$ for $\forall k\in\mathcal{K}$ in the synchronous case. Accordingly, there will be only $L_p$ VNs $\{z^m_{l},l\in\mathcal{L}_p\}$,  $L_p$ FNs $\{f^m_{\delta(l)},l\in\mathcal{L}_p\}$, and one VN $\lambda$ in Fig. \ref{PUD} and Fig. \ref{factorgraph} in the synchronous case. Then, in step 2 of the message passing algorithm, we will update the belief $b^t_\text{syn}(h^m_k)$ of VN $h^m_k$ with all the incoming messages $\{I^t_{f^{m}_{\delta(l)}\to h^m_k}(h^m_k),l\in\mathcal{L}_p\}$ and $I^t_{f^{m,k}_{h}\to h^{m}_k}(h^{m}_k)$. Furthermore, the message $I^t_{h^{m}_k\to f^{m}_{\delta(l)}}(h^{m}_k)$ in step 5 will be updated according to (\ref{VN_h_B1}), i.e. $I^t_{h^{m}_k\to f^{m}_{\delta(l)}}(h^{m}_k)\propto b^t_\text{syn}(h^m_k)/I^t_{f^{m}_{\delta(l)}\to h^m_k}(h^m_k)$. Apart from step 2 and step 5, the remaining steps of Algorithm \ref{alg:MPaSBL} will remain \emph{exactly} the same for the synchronous case. 

In the following, we will analyze the SINR at FN $f^m_{\delta(l)}$ for the synchronous case and at FN $f^{m,k}_{\delta(l)}$ for the asynchronous case. The result is presented in the following proposition.
\begin{proposition}\label{prop}
Assuming that the synchronous case and the asynchronous case share the same estimated noise precision $\hat{\lambda}_k$ and backward-passing variances $\{v^{t-1}_{h^m_{k^\prime}\to f^{m,k}_{\delta(l)}}, k^\prime\in\mathcal{K}/k\}$, the forward message from FN $f^{m,k}_{\delta(l)}$ to VN $h^m_k$ has higher SINR in the asynchronous case than in the synchronous case. 
\end{proposition}

\noindent\emph{Proof}: According to the update rule of $I^t_{f^{m,k}_{\delta(l)}\to h^m_k}(h^m_k)$ in (\ref{VN_delta_F}) and (\ref{VN_delta_F_muv}), the incoming messages $I^{t-1}_{h^m_{k^\prime}\to f^{m,k}_{\delta(l)}}(h^m_{k^\prime})$ from the other UEs $k^\prime\in\mathcal{K}/k$ are treated as interference at FN $f^{m,k}_{\delta(l)}$. Therefore, under the asynchronous case, the SINR for the message $I^t_{f^{m,k}_{\delta(l)}\to h^m_k}(h^m_k)$ can be represented as
\begin{equation}\label{aSINR}
\text{SINR}^\text{asy}_k{=}\frac{|\bar{p}^l_{k,k}|^2}{\hat{\lambda}^{-1}_k+\sum\limits_{k^\prime\neq k}v^{t-1}_{h^m_{k^\prime}\to f^{m,k}_{\delta(l)}}|\bar{p}^l_{k,k^\prime}|^2}.
\end{equation}
Similarly, under the synchronous case, the SINR for the message $I^t_{f^{m}_{\delta(l)}\to h^m_k}(h^m_k)$ can be expressed as
\begin{equation}\label{sSINR}
\text{SINR}^\text{syn}_k{=}\frac{|p^l_{k}|^2}{\hat{\lambda}^{-1}+\sum\limits_{k^\prime\neq k}v^{t-1}_{h^m_{k^\prime}\to f^{m}_{\delta(l)}}|p^l_{k^\prime}|^2},
\end{equation}
and we have $\bar{p}^l_{k,k}=p^l_{k}$ in (\ref{aSINR}) according to (\ref{eff_pilot}). Assuming that $\hat{\lambda}_k=\hat{\lambda}$ and $v^{t-1}_{h^m_{k^\prime}\to f^{m,k}_{\delta(l)}}=v^{t-1}_{h^m_{k^\prime}\to f^{m}_{\delta(l)}}$ in (\ref{aSINR}) and (\ref{sSINR}), the difference between $\text{SINR}^\text{asy}_k$ and $\text{SINR}^\text{syn}_k$ is then only determined by the difference between $|\bar{p}^l_{k,k^\prime}|^2$ and $|p^l_{k^\prime}|^2$. Without loss of any generality, take the $k^\prime$-th UE with $k^\prime<k$  as an example, where we have $\bar{p}^l_{k,k^\prime}=p^l_{k^\prime}\rho_{\tau_k-\tau_{k^\prime}}+p^{l+1}_{k^\prime}\rho_{T_s-(\tau_k-\tau_{k^\prime})}$ according to (\ref{eff_pilot}). Assuming that each pilot symbol has unit energy, we therefore have $\mathbb{E}\big(|p^l_{k^\prime}|^2\big)=1$ for $\forall k^\prime\in\mathcal{K},l\in\mathcal{L}_p$. Then, the expectation on $|p^l_{k,k^\prime}|^2$ is written as
\begin{equation}\label{genreralproof}
\begin{split}
\mathbb{E}\big(|p^l_{k,k^\prime}|^2\big)&=\rho^2_{\tau_k-\tau_{k^\prime}}+\rho^2_{T_s-(\tau_k-\tau_{k^\prime})}\\
&+2\rho_{\tau_k-\tau_{k^\prime}}\rho_{T_s-(\tau_k-\tau_{k^\prime})}\mathbb{R}\Big((p^l_{k^\prime})^*p^{l+1}_{k^\prime}\Big),
\end{split}
\end{equation}
where $\mathbb{R}(\cdot)$ takes the real part of a complex number. Since each pilot symbol has unit energy, it is easy to show that $\mathbb{E}\big(|p^l_{k,k^\prime}|^2\big)$ in (\ref{genreralproof}) is maximized when $p^l_{k^\prime}=p^{l+1}_{k^\prime}$, which is an impractical assumption for pilot design. Still, conditioned on this impractical assumption of $p^l_{k^\prime}=p^{l+1}_{k^\prime}$, we have 
\begin{equation}\label{impracticalproof}
\begin{split}
&\max\Bigg( \mathbb{E}\big(|p^l_{k,k^\prime}|^2\big) \Bigg) = \mathbb{E}\Big(|p^l_{k,k^\prime}|^2\Big|p^l_{k^\prime}=p^{l+1}_{k^\prime}\Big)\\
&=|\rho_{\tau_k-\tau_{k^\prime}}+\rho_{T_s-(\tau_k-\tau_{k^\prime})}|^2\leq 1=\mathbb{E}\big(|p^l_{k^\prime}|^2\big),
\end{split}
\end{equation}
where $|\rho_{\tau_k-\tau_{k^\prime}}+\rho_{T_s-(\tau_k-\tau_{k^\prime})}|^2= 1$ holds only for the rectangular waveform $s(t)$, which has the property $\rho_\tau+\rho_{T_s-\tau}=\rho_0$. To sum up, $\mathbb{E}\big(|p^l_{k,k^\prime}|^2\big)\leq\mathbb{E}\big(|p^l_{k^\prime}|^2\big)$, and the equality holds only under the impractical assumption that $p^l_{k^\prime}=p^{l+1}_{k^\prime}$.

The conclusion obtained from (\ref{impracticalproof}) is applicable to the general pilot design. We further consider the \emph{i.i.d} Gaussian pilot symbols, i.e. $p^l_k\sim\mathcal{CN}(0,1)$ for $\forall k\in\mathcal{K}$ and $\forall l\in\mathcal{L}_p$, which is a common choice for pilot design in message passing based detection algorithms \cite{YuWeiAMP,VBI,Sat}. In this case, we have $\mathbb{R}\Big((p^l_{k^\prime})^*p^{l+1}_{k^\prime}\Big)=0$ for (\ref{genreralproof}), and Proposition \ref{prop} can be proven with the Cauchy-Schwartz inequality,
\begin{equation}\label{proof}
\begin{split}
&\mathbb{E}\Big(|\bar{p}^l_{k,k^\prime}|^2\big|\text{Gaussian pilot}\Big)=\rho^2_{\tau_k-\tau_{k^\prime}}+\rho^2_{T_s-(\tau_k-\tau_{k^\prime})},\\
&<\int_{0}^{T_s}s^2\Big(t-(\tau_k-\tau_{k^\prime})\Big)dt\int_{0}^{T_s}s^2\Big(t\Big)dt\\
&\ \ \ + \int_{0}^{T_s}s^2\Big(t+T_s-(\tau_k-\tau_{k^\prime})\Big)dt\int_{0}^{T_s}s^2\Big(t\Big)dt=1.
\end{split}
\end{equation}
Therefore, we have $\mathbb{E}\Big(|\bar{p}^l_{k,k^\prime}|^2\big|\text{Gaussian pilot}\Big)<\mathbb{E}\big(|p^l_{k^\prime}|^2\big)$. In this way, both conclusions from (\ref{impracticalproof}) and (\ref{proof}) corroborate the claim of $\mathbb{E}\Big(\text{SINR}^\text{asy}_k\Big)>\mathbb{E}\Big(\text{SINR}^\text{syn}_k\Big)$ in Proposition \ref{prop}.
$\hfill\blacksquare$

For Gaussian pilots, according to (\ref{proof}), the factor $\rho^2_{\tau_k-\tau_{k^\prime}}+\rho^2_{T_s-(\tau_k-\tau_{k^\prime})}$ determines the multi-user interference (MUI) from the $k^\prime$-th UE to the $k$-th UE during the update of $I^t_{f^{m,k}_{\delta(l)}\to h^m_k}(h^m_k)$. Given the signal waveform $s(t)$, this MUI factor is a function of the relative delay $|\tau_k-\tau_{k^\prime}|$ between the $k^\prime$-th UE and the $k$-th UE, and we illustrate this function in Fig. \ref{waveform}(c). For two arbitrary UEs $k$ and $k^\prime$, we assume that $\tau_k$ and $\tau_{k^\prime}$ are uniformly and randomly distributed on the interval $[0,T_s]$. Then, the average of the MUI factor is further plotted as the dashed line in Fig. \ref{waveform}(c). In contrast to the asynchronous case, the synchronous case has a larger MUI factor, i.e. $\mathbb{E}\big(|p^l_{k^\prime}|^2\big)=1$. As a result, the message $I^t_{f^{m,k}_{\delta(l)}\to h^m_k}(h^m_k)$ under the asynchronous case will be more reliable than the message $I^t_{f^{m}_{\delta(l)}\to h^m_k}(h^m_k)$ under the synchronous case. Furthermore, according to (\ref{belief}) and (\ref{belief_muv}), the belief of the VN $h^m_k$ will also be more reliable under the asynchronous case. In summary, the proposed method exploits asynchronization between different UEs to suppress MUI, leading to improved detection accuracy.
\subsection{Asynchronous Case with Imperfect Delay Information}\label{nonidealTA}
So far, we have derived and analyzed the PUDMP-SBL-aUADCE algorithm based on the assumption that the BS has \emph{perfect} knowledge of the delay information $\{\tau_k,k\in\mathcal{K}\}$. In this subsection, we consider a more practical scenario with imperfect delay information, which may be potentially caused by device mobility or timing error of the low-cost oscillators at the devices. Specifically, assume that the $k$-th UE is activated, and denote $\tilde{\tau}_k$ as its \emph{actual} transmission delay in this round of GF-RA. On the other hand, the BS only knows the imperfect delay information $\tau_k$, and the delay error is denoted as $\tilde{\tau}^\text{err}_k\overset{\Delta}{=}\tilde{\tau}_k-\tau_k$. In the following, we will analyze the impacts of the imperfect delay information, as well as the benefits of adopting the SBL method in this case.

Similar to (\ref{time_rcv}), denote $\tilde{y}^m(t)$ as the received baseband waveform signal with actual transmission delay $\{\tilde{\tau}_k,k\in\mathcal{K}\}$, and we have
\begin{equation}\label{time_rcv_actual}
\tilde{y}^m(t)=\sum\limits^{K}_{k=1}\sum\limits^{L_p}_{l=1}\alpha_kg^m_kp^l_ks\Big(t-(l-1)T_s-\tilde{\tau}_k\Big) + \tilde{n}(t).
\end{equation}
Then, the BS will perform the matched filtering and sampling operations on $\tilde{y}^m(t)$, using the imperfect delay information $\{\tau_k,k\in\mathcal{K}\}$. In this case, similar to (\ref{digital_rcv}), the $l$-th sampled output $\tilde{y}^m_k(l)$ of the $k$-th matched filter on the $m$-th antenna is written as
\begin{equation}\label{digital_rcv_Imp}
\begin{split}
&\tilde{y}^m_k(l)=\int^{lT_s+\tau_k}_{(l-1)T_s+\tau_k}\tilde{y}^m(t)s\Big(t-(l-1)T_s-\tau_k\Big)dt,\\
&\overset{(e)}{=}\rho_{|\tilde{\tau}^\text{err}_k|}h^m_kp^l_k+\rho_{T_s-|\tilde{\tau}^\text{err}_k|}h^m_kp^{l-\text{sign}(\tilde{\tau}^\text{err}_k)}_k+\tilde{n}^m_k(l)\\
&\ \ \ +\sum\limits_{\{k^\prime,2T_s\geq\tilde{\tau}_{k^\prime}-\tau_k>T_s\}}\!\!\!\!\!\!\!\!\!\!\!\!\!h^m_{k^\prime}\Big(p^{l-2}_{k^\prime}\rho_{2T_s+\tau_k-\tilde{\tau}_{k^\prime}}+p^{l-1}_{k^\prime}\rho_{\tilde{\tau}_{k^\prime}-\tau_k-T_s}\Big)\\
&\ \ \ +\sum\limits_{\{k^\prime,T_s\geq\tilde{\tau}_{k^\prime}-\tau_k>0\}}\!\!\!\!\!\!\!\!h^m_{k^\prime}\Big(p^{l-1}_{k^\prime}\rho_{T_s+\tau_k-\tilde{\tau}_{k^\prime}}+p^{l}_{k^\prime}\rho_{\tilde{\tau}_{k^\prime}-\tau_k}\Big)\\
&\ \ \ +\sum\limits_{\{k^\prime,T_s>\tau_k-\tilde{\tau}_{k^\prime}\geq 0\}}\!\!\!\!\!\!\!\!h^m_{k^\prime}\Big(p^{l}_{k^\prime}\rho_{\tau_k-\tilde{\tau}_{k^\prime}}+p^{l+1}_{k^\prime}\rho_{T_s+\tilde{\tau}_{k^\prime}-\tau_k}\Big)\\
&\ \ \ +\sum\limits_{\{k^\prime,2T_s>\tau_k-\tilde{\tau}_{k^\prime}\geq T_s\}}\!\!\!\!\!\!\!\!\!\!\!\!\!h^m_{k^\prime}\Big(p^{l+1}_{k^\prime}\rho_{\tau_k-\tilde{\tau}_{k^\prime}-T_s}+p^{l+2}_{k^\prime}\rho_{2T_s+\tilde{\tau}_{k^\prime}-\tau_k}\Big),\\
&\overset{(f)}{=}\rho_{|\tilde{\tau}^\text{err}_k|}h^m_kp^l_k+\rho_{T_s-|\tilde{\tau}^\text{err}_k|}h^m_kp^{l-\text{sign}(\tilde{\tau}^\text{err}_k)}_k+\tilde{n}^m_k(l)+\tilde{\Theta}^m_k(l),\\
&\overset{(g)}{=}\rho_{|\tilde{\tau}^\text{err}_k|}h^m_kp^l_k+\sum\limits_{k^\prime\neq k}h^m_{k^\prime}\bar{p}^l_{k,k^\prime}+\tilde{w}^m_k(l),\\
\end{split}
\end{equation}
where $h^m_k=\alpha_kg^m_k$, $\tilde{n}^m_k(l)$ is the AWGN, equation ($e$) in (\ref{digital_rcv_Imp}) is derived based on the assumption that $|\tilde{\tau}^\text{err}_k|\leq \frac{1}{2}T_s$ for $\forall k\in\mathcal{K}$, and the sign function is $\text{sign}(x)=-1,0,1$ for $x<0$, $x=0$, and $x>0$, respectively. With some straightforward derivation, we can readily extend equation ($e$) in (\ref{digital_rcv_Imp}) to a more general case with $|\tilde{\tau}^\text{err}_k|>\frac{1}{2}T_s$. However, we will show that $|\tilde{\tau}^\text{err}_k|>\frac{1}{2}T_s$ indicates a serious mismatch problem between the known delay information $\tau_k$ and the actual delay $\tilde{\tau}_k$, which causes detection failure for the $k$-th UE.

To start with, we consider the first term on the right hand side (RHS) of equation ($e$) in (\ref{digital_rcv_Imp}). In the first term, the same unknown coefficient $\rho_{|\tilde{\tau}^\text{err}_k|}$ affects the VNs $\{h^m_k,m\in\mathcal{M}\}$ of the $k$-th UE on all the $M$ antennas. Note that if the $k$-th UE is active, we have $h^m_k\sim\mathcal{CN}(0,1)$ and $\rho_{|\tilde{\tau}^\text{err}_k|}h^m_k\sim\mathcal{CN}(0,\rho^2_{|\tilde{\tau}^\text{err}_k|})$ for $\forall m\in\mathcal{M}$. On the other hand, we adopt the conditional prior distribution $p(h^m_k|\gamma_k)=\mathcal{CN}(h^m_k|0,\gamma^{-1}_k)$ with $p(\gamma_k)=Gam(\gamma_k;\epsilon,\eta)$ in the SBL method. Both parameters $\gamma_k$ and $\epsilon$ are iteratively updated in the proposed PUDMP-SBL-aUADCE algorithm. In other words, the PUDMP-SBL-aUADCE algorithm can implicitly learn the unknown power factor $\rho^2_{|\tilde{\tau}^\text{err}_k|}$ caused by delay error via estimating parameter $\gamma_k$. In addition, as pointed out in Remark \ref{rmk}, the shape parameter $\epsilon$ functions as a selective amplifier for the VNs $\{\gamma_k,k\in\mathcal{K}\}$ \cite{Justify_epsilon}, and updating $\epsilon$ benefits the robustness of the PUDMP-SBL-aUADCE algorithm against unknown delay error $\{\tilde{\tau}^\text{err}_k,k\in\mathcal{K}\}$. 

Then, we consider the second term $\rho_{T_s-|\tilde{\tau}^\text{err}_k|}h^m_kp^{l-\text{sign}(\tilde{\tau}^\text{err}_k)}_k$ on the RHS of equation ($e$) in (\ref{digital_rcv_Imp}). Note that even if the factor $\rho_{|\tilde{\tau}^\text{err}_k|}$ is accurately learned from the PUDMP-SBL-aUADCE algorithm, it is still difficult to detect $\text{sign}(\tilde{\tau}^\text{err}_k)$ and $p^{l-\text{sign}(\tilde{\tau}^\text{err}_k)}_k$. Therefore, a practical yet effective solution is to treat the second term $\rho_{T_s-|\tilde{\tau}^\text{err}_k|}h^m_kp^{l-\text{sign}(\tilde{\tau}^\text{err}_k)}_k$ as a noise term. Furthermore, if $|\tilde{\tau}^\text{err}_k|>\frac{1}{2}T_s$, the power of this noise term $\rho_{T_s-|\tilde{\tau}^\text{err}_k|}h^m_kp^{l-\text{sign}(\tilde{\tau}^\text{err}_k)}_k$ will exceed the power of the first term $\rho_{|\tilde{\tau}^\text{err}_k|}h^m_kp^l_k$, which may lead to failed estimation of $\rho_{|\tilde{\tau}^\text{err}_k|}$ and $h^m_k$. In other words, if $|\tilde{\tau}^\text{err}_k|>\frac{1}{2}T_s$, the matched filter aligned with imperfect delay $\tau_k$ will fail to match the target pilot symbol $p^l_k$, which seriously undermines the accuracy of the PUDMP-SBL-aUADCE algorithm. 

Finally, we consider the last four accumulative-addition terms on the RHS of equation ($e$). For notation convenience, the sum of these four terms is denoted as $\tilde{\Theta}^m_k(l)$ in equation ($f$) of (\ref{digital_rcv_Imp}), and we will have $\tilde{\Theta}^m_k(l)=\sum\limits_{k^\prime\neq k}h^m_{k^\prime}\bar{p}^l_{k,k^\prime}$ if $\tilde{\tau}^\text{err}_{k^\prime}=0$ for $\forall k^\prime\in\mathcal{K}/k$. However, due to the existence of the random delay error $\{\tilde{\tau}^\text{err}_k,k\in\mathcal{K}\}$, it is difficult to determine $\tilde{\Theta}^m_k(l)$. Again, we treat the difference between $\tilde{\Theta}^m_k(l)$ and $\sum\limits_{k^\prime\neq k}h^m_{k^\prime}\bar{p}^l_{k,k^\prime}$ as an equivalent noise. In this way, we can write $\tilde{y}^m_k(l)$ as in equation ($g$) of (\ref{digital_rcv_Imp}), where $\tilde{w}^m_k(l)$ collects all the \emph{equivalent} noise terms, i.e. $\tilde{w}^m_k(l)=\rho_{T_s-|\tilde{\tau}^\text{err}_k|}h^m_kp^{l-\text{sign}(\tilde{\tau}^\text{err}_k)}_k+\tilde{n}^m_k(l)+\big(\tilde{\Theta}^m_k(l)-\sum\limits_{k^\prime\neq k}h^m_{k^\prime}\bar{p}^l_{k,k^\prime}\big)$. In this way, equation ($g$) is employed by the proposed PUDMP-SBL-aUADCE algorithm for UAD and CE, where the noise power of $\tilde{w}^m_k(l)$ is estimated by updating $\hat{\lambda}_k$ in (\ref{E_Lambda}).
\section{Simulation Results}\label{simu_section}
\begin{figure*}
	\centering
	\includegraphics[width=1.7\columnwidth]{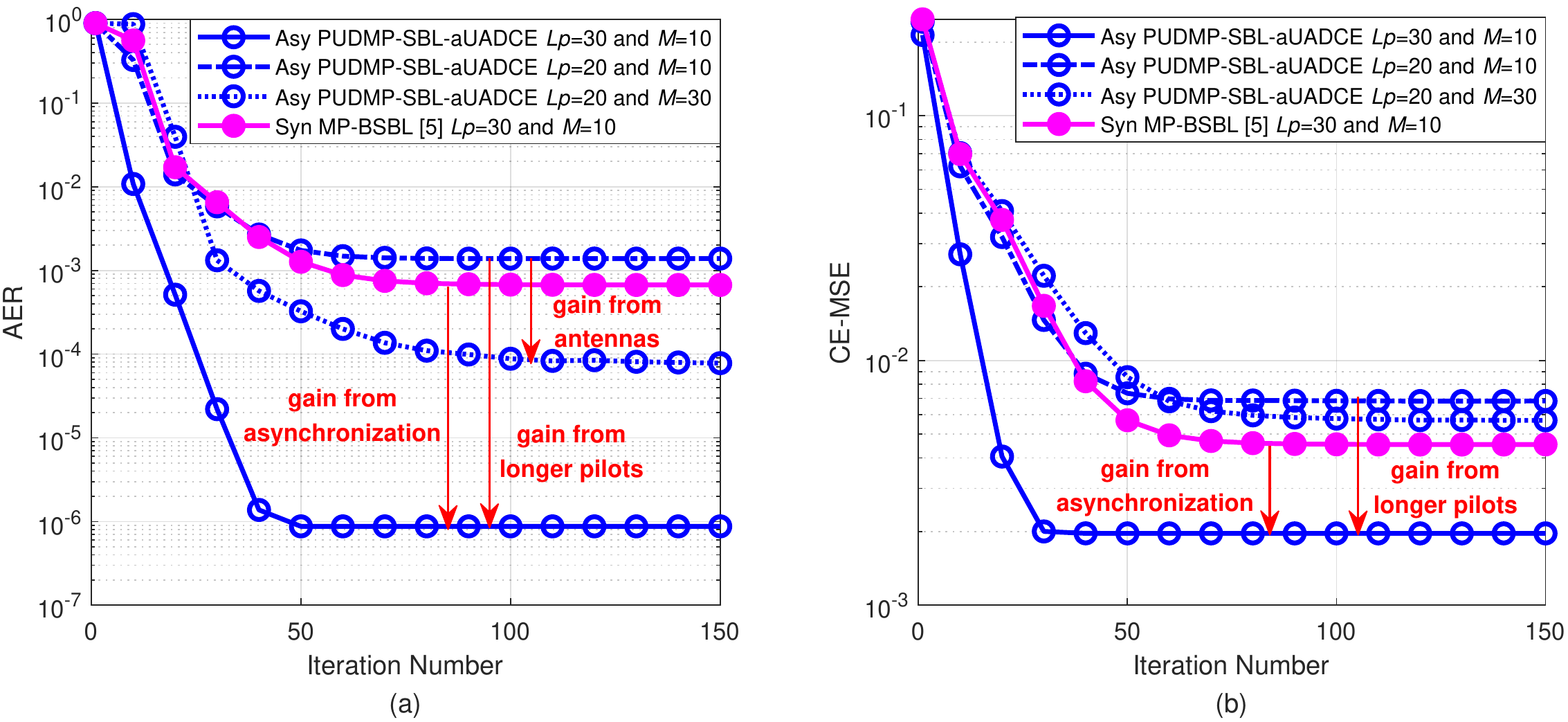}\vspace{-0.2cm}
	\caption{Convergence performance on (a) AER and (b) CE-MSE with $K=200$, $p_a=0.1$, and $\text{SNR}=10 \text{dB}$.}
	\label{convergence}
\end{figure*}
\begin{figure*}
	\centering
	\includegraphics[width=1.7\columnwidth]{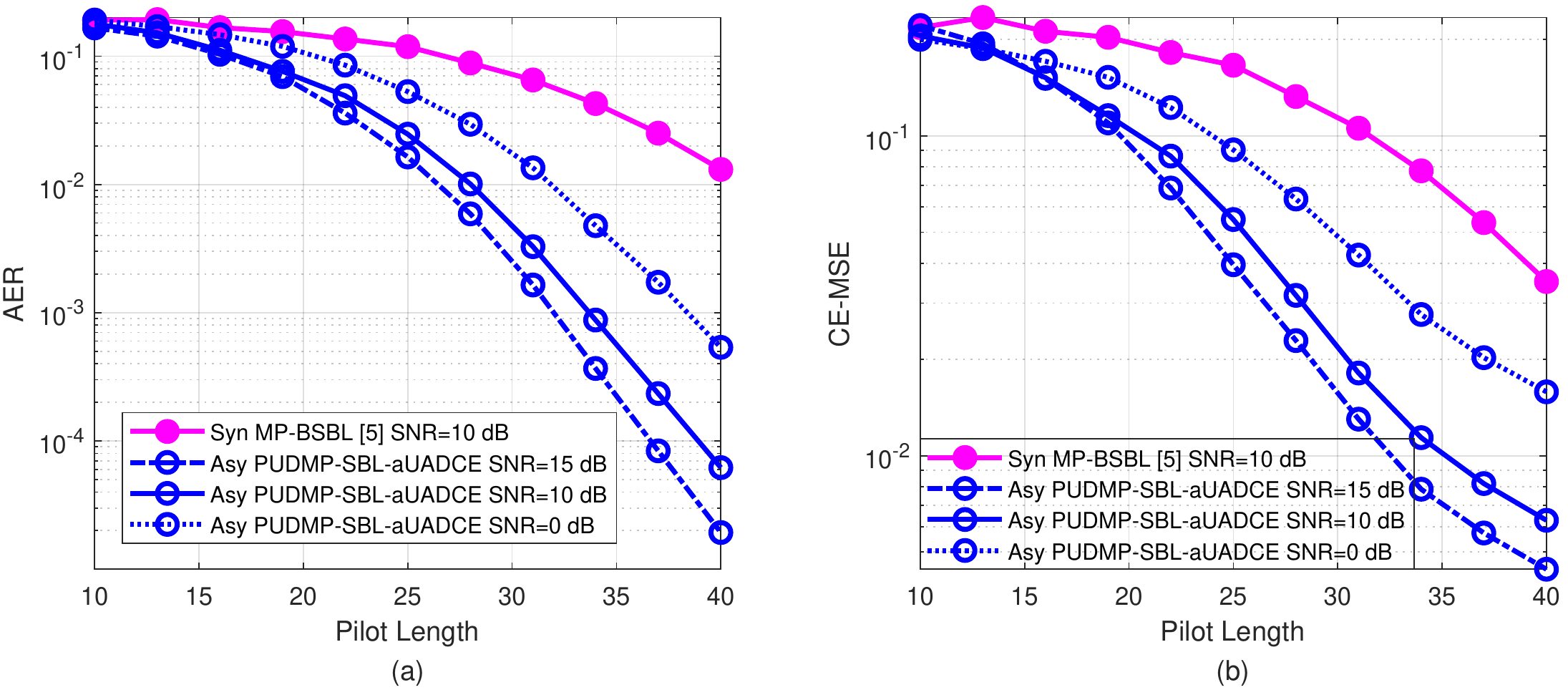}\vspace{-0.2cm}
	\caption{Impacts of pilot length $L_p$ on (a) AER and (b) CE-MSE performance with $K=200$, $M=10$, $p_a=0.2$, and $N_{it}=80$.}
	\label{Lp}
\end{figure*}
In this section, extensive simulations are conducted to examine the performances of the PUDMP-SBL-aUADCE algorithm. Specifically, we consider the UAD and CE accuracy, and define the activity-detection error rate (AER) and channel-estimation mean square error (CE-MSE) as performance metrics, i.e., $\text{AER}={\sum^K_{k=1}|\alpha_{k}-\hat{\alpha}_{k}|}/{K}$ and $\text{CE-MSE}={\sum^K_{k=1}\sum^M_{m=1}|h^m_k-\hat{h}^m_{k}|^2}/{KM}$. Through the following simulations, we investigate the convergence performance, and the impacts of different system parameters, i.e. the pilot length $L_p$, the antenna number $M$, the activation probability $p_a$, the delay error $\tilde{\tau}^\text{err}$, and the signal-to-noise ratio (SNR) which is defined as $\text{SNR}=10\log_{10}\frac{1}{\sigma^2_n}$. For comparison, we also simulate the performances of the unitary AMP based SBL (UAMP-SBL) algorithm \cite{Firts_UAMP,Justify_epsilon}, the block orthogonal matching pursuit (BOMP) algorithm \cite{BOMP}, the orthogonal AMP (OAMP) algorithm \cite{OAMP}, the genie-aided minimum mean square error (GA-MMSE) algorithm, as well as the MP-BSBL algorithm \cite{ZYYMP} which can be considered as a synchronous version of the PUDMP-SBL-aUADCE algorithm. Here, the GA-MMSE algorithm is aided with ideal user activity information, and therefore it provides the CE-MSE lower bound. We will further show that the proposed PUDMP-SBL-aUADCE algorithm can closely approach this lower bound within a wide range of SNR.
\begin{figure*}
	\centering
	\includegraphics[width=1.7\columnwidth]{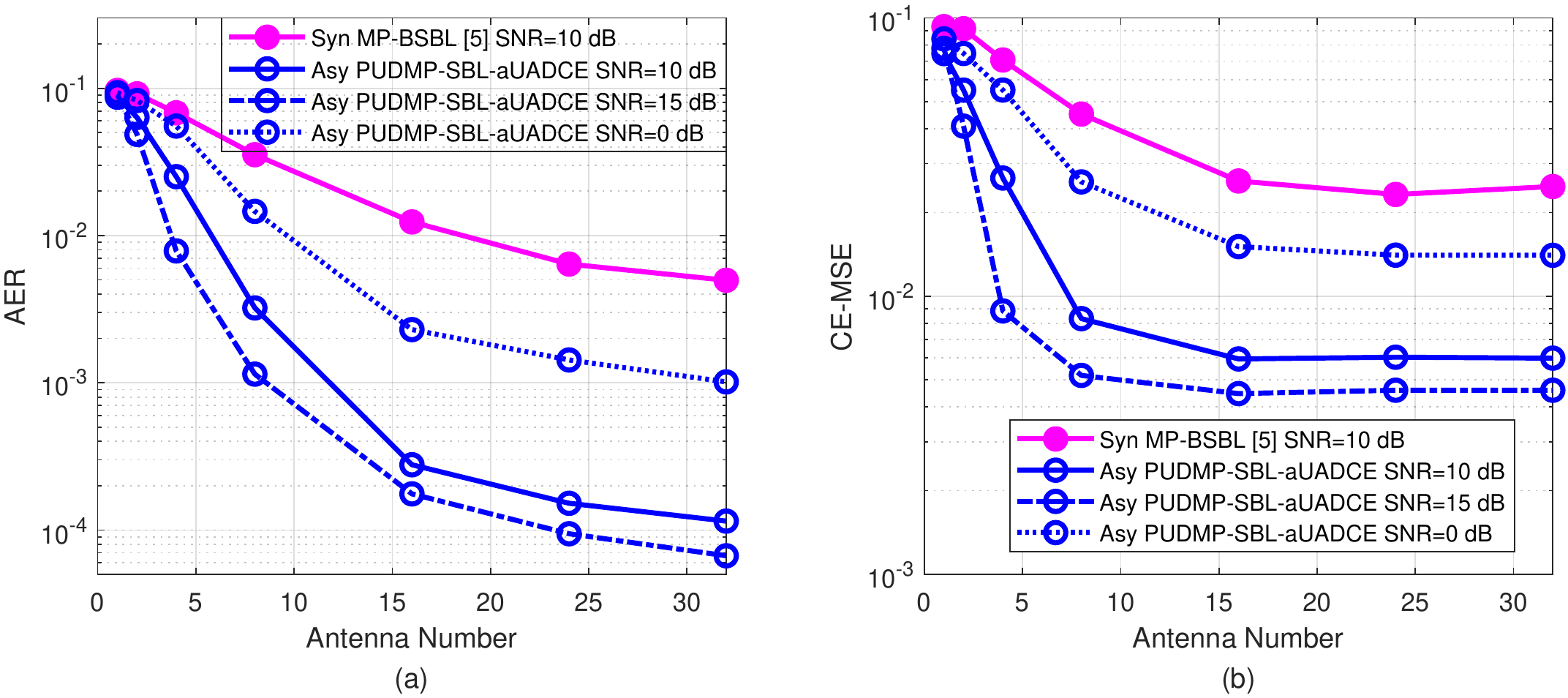}
	\caption{Impacts of antenna number $M$ on (a) AER and (b) CE-MSE performance with $K=200$, $Lp=20$, $p_a=0.1$, and $N_{it}=80$.}
	\label{M}
\end{figure*} 
\begin{figure*}
\centering
\includegraphics[width=1.7\columnwidth]{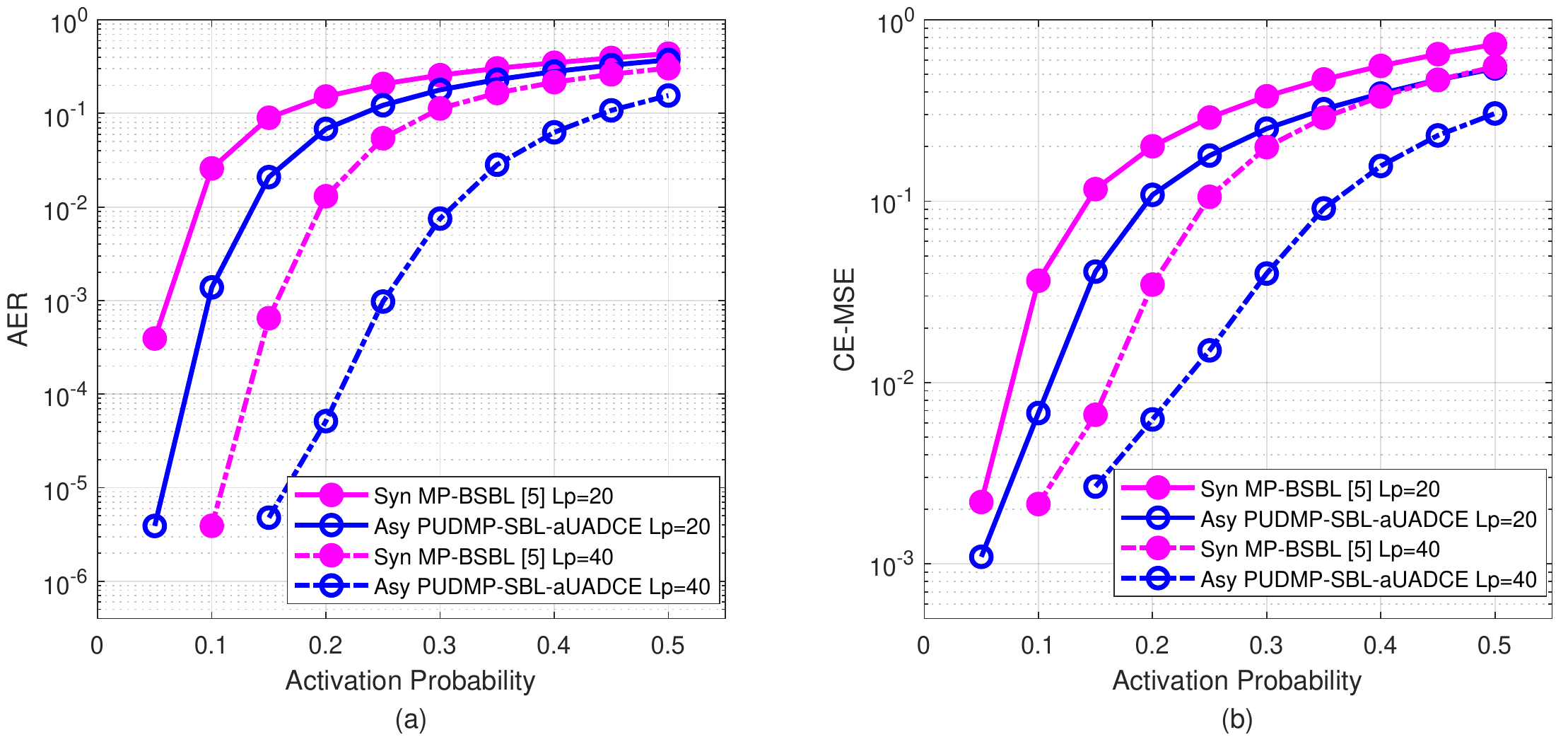}
\caption{Impacts of activation probability $p_a$ on (a) AER and (b) CE-MSE performance with $K=200$, $M=10$, $\text{SNR}=10 \text{dB}$, and $N_{it}=80$.}
\label{Pa}
\end{figure*}
\subsection{Convergence Performance}\label{cov}
To start with, we simulate the convergence performance of the PUDMP-SBL-aUADCE algorithm with different iteration number $N_{it}$. The AER and CE-MSE results are illustrated in Fig. \ref{convergence}(a) and Fig. \ref{convergence}(b), respectively. It is shown that with the same pilot length $L_p=30$ and antenna number $M=10$, the PUDMP-SBL-aUADCE algorithm in the asynchronous case could significantly outperform the MP-BSBL algorithm in the synchronous case, especially in terms of the AER performance. This performance gain from asynchronization is consistent with the conclusion from Proposition \ref{prop}. Consider only the asynchronous case and fix $M=10$, it is shown that increasing $L_p$ from $20$ to $30$ could improve the convergence speed of the PUDMP-SBL-aUADCE algorithm, as well as the UAD and CE accuracy. Fixing $L_p=20$ and increasing $M$ from $10$ to $30$ for the PUDMP-SBL-aUADCE algorithm, we can also observe a performance gain for the AER, while the CE-MSE performance will not get improved with larger $M$. This observation can be explained from the fact that the channel gains $\{h^m_k,m\in\mathcal{M}\}$ for each active UE $K$ are \emph{i.i.d.} on different antennas. In this case, as long as the UAD accuracy is sufficiently high, increasing the antenna number $M$ will not lead to higher CE accuracy. 
\subsection{Impacts of Pilot Length}
As shown in Fig. \ref{convergence}, the pilot length has remarkable impacts on the convergence and accuracy of the PUDMP-SBL-aUADCE algorithm, and simulation results with different pilot length are plotted in Fig. \ref{Lp}. We can also observe the asynchronization gain from Fig. \ref{Lp}, i.e. with the same $\text{SNR}=10 \text{dB}$, the PUDMP-SBL-aUADCE algorithm in the asynchronous case still outperforms the MP-BSBL algorithm in the synchronous case. For example, fix $\text{SNR}=10 \text{dB}$ and $L_p=40$, the AER of the PUDMP-SBL-aUADCE algorithm is lower than that of the MP-BSBL algorithm by \emph{more than two magnitude orders}. In addition, it is shown that increasing the pilot length could also effectively improve the CE-MSE performances. 
\begin{figure*}
	\centering
	\includegraphics[width=1.7\columnwidth]{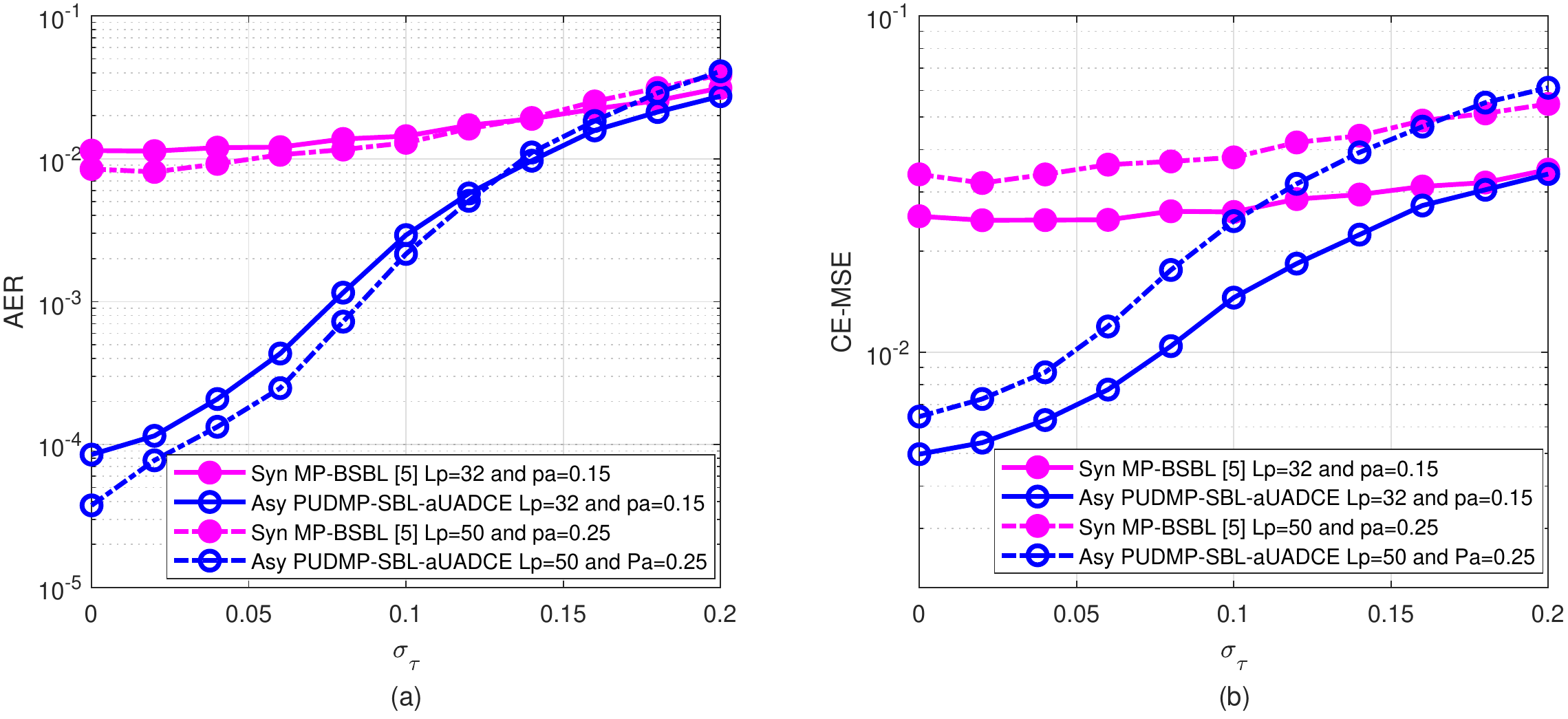}
	\caption{Impacts of $\sigma_\tau$  on (a) AER and (b) CE-MSE performance with $K=200$, $M=10$, $\text{SNR}=10 \text{dB}$, and $N_{it}=80$.}
	\label{sigmatau}
\end{figure*} 
\subsection{Impacts of Antenna Number}
The detection performances with different antenna number $M$ are shown in Fig. \ref{M}. Again, we can still observe the asynchronization gain when SNR is fixed as 10 dB. Furthermore, it is shown that increasing the antenna number $M$ can always improve the UAD detection accuracy. This observation is explained by the fact that with more antennas, more messages $\{\mu^t_{h^{m}_k},v^t_{h^{m}_k},m\in\mathcal{M}\}$ will be employed in (\ref{Gam_up}) to update the activity-related hyper-parameter $\hat{\gamma}_k$ for each UE $k$. However, the CE-MSE performance converges at $M=16$, and further increasing $M$ will not improve the CE accuracy. Explanation for this observation can be found in Section \ref{cov}. 
\subsection{Impacts of Activation Probability}
The impacts of the activation probability $p_a$ on the PUDMP-SBL-aUADCE algorithm are illustrated in Fig. \ref{Pa}. It is shown that, when the random access scenario is crowded with more active UEs, both the AER and the CE-MSE performances will inevitably deteriorate due to the increased MUI. As explained in Proposition \ref{prop}, the PUDMP-SBL-aUADCE algorithm can effectively mitigate the MUI in the asynchronous case. Therefore, given the same pilot length $L_p$, the PUDMP-SBL-aUADCE algorithm can always outperform the MP-BSBL algorithm within a wide range of $p_a$. Especially, in the low $p_a$ regime, the PUDMP-SBL-aUADCE algorithm can improve the AER of the MP-BSBL algorithm by \emph{more than two magnitude orders}, which makes the PUDMP-SBL-aUADCE algorithm particularly favorable to mMTC applications that are featured by sparsely activated devices.
\subsection{Impacts of Imperfect Delay Information}
We consider a more practical scenario with imperfect delay information potentially caused by device mobility or timing error. Recall that the delay error $\tilde{\tau}^\text{err}_k$ is defined as the difference between the actual transmission delay $\tilde{\tau}_k$ and the imperfect delay information $\tau_k$ at the BS, i.e. $\tilde{\tau}^\text{err}_k=\tilde{\tau}_k-\tau_k$. Without loss of any generality, we assume that $\tilde{\tau}^\text{err}_k\sim\mathcal{N}\big(0,(\sigma_\tau T_s)^2\big)$ for $\forall k\in\mathcal{K}$. Note that the assumption of Gaussian-distributed $\tilde{\tau}^\text{err}_k$ is adopted simply for simulation convenience, so that the severity of the delay error can be controlled with one parameter $\sigma_\tau$. The performances of the PUDMP-SBL-aUADCE algorithm under different $\sigma_\tau$ are illustrated in Fig. \ref{sigmatau}. It is shown that, even in the presence of random delay error $\tilde{\tau}^\text{err}_k$, the PUDMP-SBL-aUADCE algorithm still exhibits higher UAD and CE accuracy than the MP-BSBL algorithm. In addition, it is shown that both algorithms will fail to work when $\sigma_\tau$ approaches $0.2$. This observation can be explained by the fact that a larger $\sigma_\tau$ is more likely to incur the mismatch problem with $|\tilde{\tau}^\text{err}_k|>\frac{1}{2}T_s$. As explained in Section \ref{nonidealTA}, if $|\tilde{\tau}^\text{err}_k|>\frac{1}{2}T_s$, the matched filter aligned with imperfect delay $\tau_k$ will fail to match the target pilot symbol, which eventually leads to failed UAD and CE. 
\begin{figure}
	\centering
	\includegraphics[width=0.9\columnwidth]{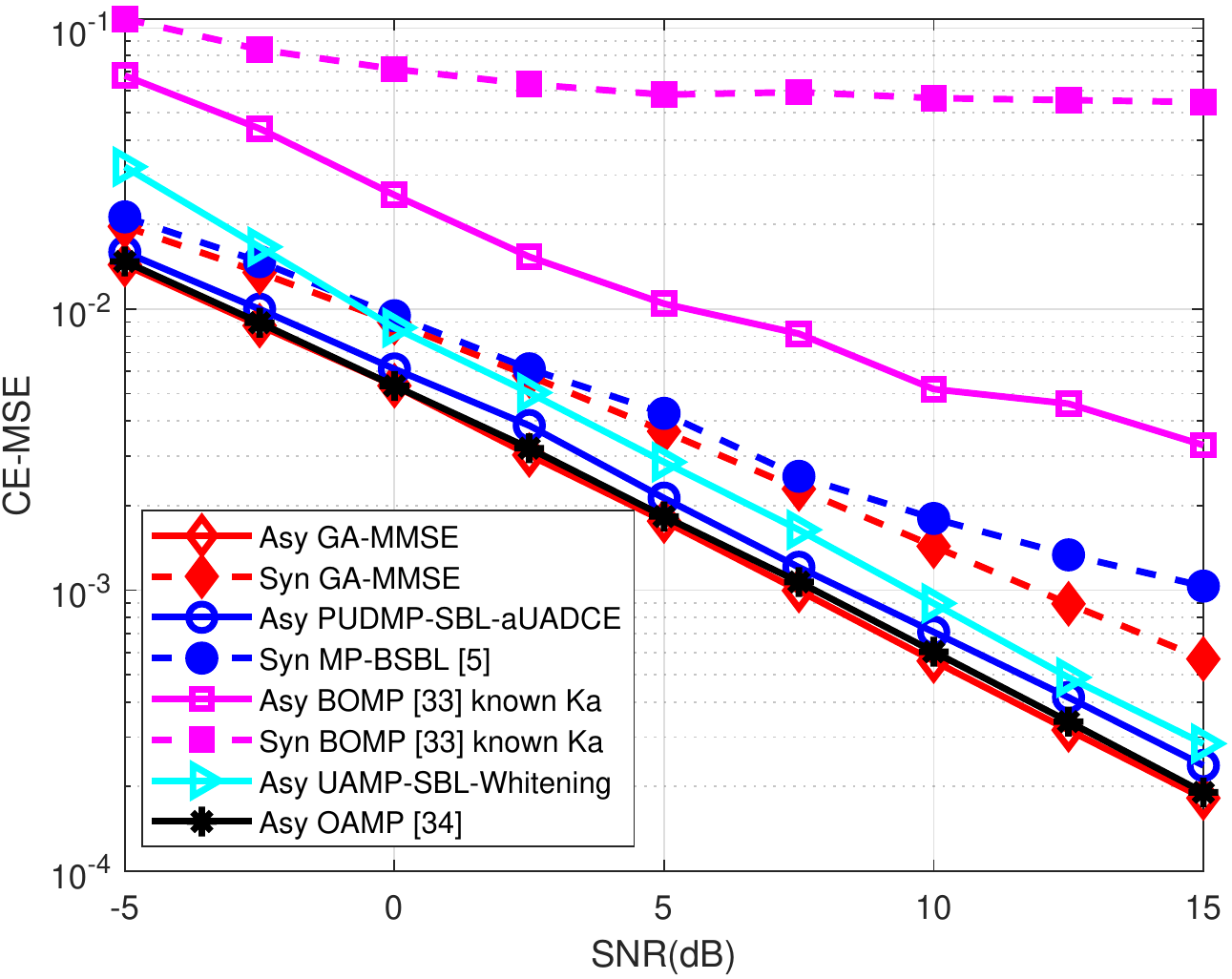}
	\caption{CE-MSE performance under different $\text{SNR}$ with $K=200$, $M=32$, $L_p=30$, and $p_a=0.1$. Solid lines represent the performances of different algorithms in the asynchronous case, while dashed lines represent the performances of different algorithms in the synchronous case.}
	\label{snr}
\end{figure} 
\subsection{Performance Comparison under Different SNR}
For comparison convenience, all the sampled outputs on different antennas are re-organized in the linear mixing model,
\begin{equation}\label{tot}
\bar{\bm{Y}}=\bar{\bm{P}}\bm{H}+\bar{\bm{N}},
\end{equation}
where $\bar{\bm{Y}}=[\bm{Y}_1^T,\ldots,\bm{Y}_k^T,\ldots,\bm{Y}_K^T]^T$ has the size $KL_p\times M$, $\bm{Y}_k=[\bm{y}^1_k,\ldots,\bm{y}^m_k,\ldots,\bm{y}^M_k]$ has the size $L_p\times M$, and $\bm{Y}_k$ collects the sampled outputs from the $k$-th matched filter on all the $M$ antennas, while $\bm{y}^m_k$ is given in (\ref{rcv_vec}). $\bar{\bm{P}}=[\bar{\bm{P}}^T_1,\ldots,\bar{\bm{P}}^T_k,\ldots,\bar{\bm{P}}^T_K]^T$, and the element $\bar{p}^l_{k,k^\prime}$ on the $l$-th row and $k^\prime$-th column of $\bar{\bm{P}}_k$ is given in (\ref{eff_pilot}). $\bm{H}$ is the effective channel matrix explained in Section \ref{probform}. Each column $\bar{\bm{n}}^m,m\in\mathcal{M}$ of $\bar{\bm{N}}$ is distributed as $\bar{\bm{n}}^m\sim\mathcal{CN}(\bm{0},\bar{\bm{\Sigma}})$ with the noise covariance matrix $\bar{\bm{\Sigma}}$ written as $\bar{\bm{\Sigma}}=\bm{\Sigma}\sigma^2_n$. Denote $\bm{\Sigma}_{i,j}$ as the element on the $i$-th row and $j$-th column of $\bm{\Sigma}$. According to (\ref{cor_noise_sample}), $\bm{\Sigma}$ is a symmetric matrix with $\bm{\Sigma}_{(k-1)L_p+l,(k^\prime-1)L_p+l}=\rho_{\tau_k-\tau_{k^\prime}}$ and $\bm{\Sigma}_{(k-1)L_p+l,(k^\prime-1)L_p+l+1}=\rho_{T_s-(\tau_k-\tau_{k^\prime})}$ for $\forall k^\prime\leq k$. In this way, we can adopt the GA-MMSE algorithm, a modified version of the UAMP-SBL algorithm \cite{Justify_epsilon}, the BOMP algorithm \cite{BOMP}, and the OAMP algorithm \cite{OAMP} for comparison. The active-UE number $K_a$ is assumed known to the BOMP algorithm, which provides an \emph{ideal stopping criterion} for the matching pursuit procedure in the BOMP algorithm. As the UAMP-SBL algorithm developed in \cite{Justify_epsilon} is for white noise, a noise-whitening operation is incorporated by left-multiplying $\bm{\Sigma}^{-1/2}$ on both sides of (\ref{tot}), which is called as UAMP-SBL-Whitening. In addition, the GA-MMSE algorithm is aided with ideal user activity information, so that it provides the CE-MSE lower bound. The simulation results are illustrated in Fig. \ref{snr}.

It is shown in Fig. \ref{snr} that, in the asynchronous case, both the GA-MMSE algorithm and the BOMP algorithm outperform their counterparts in the synchronous case. Similarly, the PUDMP-SBL-aUADCE algorithm also outperforms the MP-BSBL algorithm. The UAMP-SBL-Whitening algorithm also outperforms the algorithms in the synchronous case. However, the computational complexity of this noise-whitening operation is $\mathcal{O}(K^3L_p^3)$. In contrast, the proposed receiver design employs the PUD factor graph with related message update rules to address this sample correlation problem. It is shown that the proposed PUDMP-SBL-aUADCE algorithm can closely approach the CE-MSE lower bound within a wide range of SNR. Similarly, the correlation feature of the noise samples in $\bar{\bm{N}}$ is also exploited by the Linear MMSE (LMMSE) module of the OAMP algorithm, so that the OAMP algorithm also exhibits the bound-approaching performances. However, this LMMSE module incurs $\mathcal{O}(K^3L^3_p)$ computations for the OAMP algorithm, which is prohibitively high for mMTC applications with a massive UE number $K$. Recall from Section \ref{complex} that the computations of the PUDMP-SBL-aUADCE algorithm is inherently feasible for parallel implementations. Therefore, the PUDMP-SBL-aUADCE algorithm serves as a bound-approaching yet practical solution for the joint UAD and CE problem in asynchronous GF-RA.
\section{Conclusions}\label{conclusion}
In this paper, a PUDMP-SBL-aUADCE algorithm was proposed to address the joint UAD and CE problem for asynchronous GF-RA. Theoretical analysis shows that the PUDMP-SBL-aUADCE algorithm provides higher SINR in the asynchronous case than in the synchronous case. Considering the potential timing error from the low-cost UEs, the impacts of imperfect delay profile were investigated, revealing the advantages of adopting the SBL method in this case. Finally, extensive simulation results were provided to demonstrate the superior performances of the PUDMP-SBL-aUADCE algorithm.
\ifCLASSOPTIONcaptionsoff
  \newpage
\fi



%

\end{document}